\documentclass[useAMS,usenatbib,onecolumn]{mnras}
\usepackage{amsmath,bm}
\usepackage{dcolumn}
\usepackage[utf8]{inputenc}
\usepackage{graphics,graphicx}
\usepackage{float}
\usepackage{threeparttable}
\usepackage{url}
\usepackage{epstopdf}

\usepackage{longtable}

\usepackage{color}

\newcommand{\bra}[1]{\langle #1|}
\newcommand{\ket}[1]{|#1\rangle}

\newcommand{\COtwo}{$^{12}$C$^{16}$O$_2$}
\newcommand{\2}{$_2$}

\newcommand{\Cv}[1]{${\mathcal C}_{#1{\rm v}}$}

\newcommand{\cm}{cm$^{-1}$}
\newcommand{\ai}{\textit{ab initio}}
\newcommand{\um}{$\mu$m}

\newcommand{\rme}{_{\rm e}}

\graphicspath{{eps/}}

\title[ExoMol line lists -- XXXIX. CO$_2$]{ExoMol line lists -- XXXIX. Ro-vibrational molecular line list for CO$_2$}

\date{\today}

\author[]{S. N. Yurchenko$^1$, Thomas M. Mellor$^1$, Richard S. Freedman$^{2,3}$  and J. Tennyson$^1$\thanks{The corresponding author: j.tennyson@ucl.ac.uk} \vspace*{4mm}\\
$^1$Department of Physics and Astronomy, University College London, Gower Street, WC1E 6BT London, UK; \\
 $^2$NASA Ames Research Center, Mail Stop 245-3, Moffett Field, CA 94035-1000, USA; \\
$^3$ SETI Institute, Mountain View, CA, USA.
}

\begin{document}

\label{firstpage}

\maketitle

\begin{abstract}

%2 557 549 946 total and 3 480 477

A new hot line list for the main isotopologue of CO$_2$, $^{12}$C$^{16}$O$_2$ is presented. The line list consists of almost 2.5 billion  transitions between 3.5 million rotation-vibration states of CO$_2$ in its ground electronic state, covering the wavenumber range 0--20~000~cm$^{-1}$  ($\lambda >0.5$~\um) with the upper and lower energy thresholds of 36~000~cm$^{-1}$ and  16~000~cm$^{-1}$, respectively. The ro-vibrational energies and wavefunctions are computed variationally using the Ames-2 accurate empirical potential energy surface. The ro-vibrational transition probabilities in the form of Einstein coefficients are computed using an accurate \textit{ab initio} dipole moment surface using variational program TROVE. A new implementation of TROVE which uses an exact nuclear-motion kinetic energy operator is employed. Comparisons with the existing hot line lists are presented. The line list should be useful for atmospheric retrievals of exoplanets and cool stars. The UCL-4000 line list is available from the CDS and ExoMol databases.

\end{abstract}


\section{Introduction}

Carbon dioxide is well-known and much studied constituent
of the Earth's atmosphere. However, it is also an important
constituent of planetary atmospheres. The Venusian atmosphere is 95\%\ CO$_2$ which therefore
dominates its opacity  \citep{14SnStGr.CO2}. Similarly studies
of exoplanets have emphasised the importance of CO$_2$. It was one
of the first molecules detected in the atmospheres of hot Jupiter
exoplanets \citep{09SwTiVa.CO2,09SwVaTi.CO2,sdg10} where it
provides an important measure of the C/O ratio on the planet
\citep{13MoKaVi}. Similarly it is considered an important
marker in directly imaged exoplanets \citep{13MoKaVi} and
the atmospheres of lower mass planets are expected to be
dominated by water and CO$_2$ \citep{16MaHaTi}. \citet{16HeLyxx.CO2}
provide a comprehensive study of CO$_2$   abundances in exoplanets.
Recently,
\citet{20BaCrGu.CO2} detected emission and absorption from excited vibrational bands of CO$_2$
in the mid-infrared spectra of the M-type Mira variable R Tri using the Spitzer
infrared spectrograph (IRS).

Most of the environments discussed above are considerably  hotter
then the Earth: for example Venus is at about 735~K and hot Jupiter
exoplanets have typical atmospheric temperatures in excess of 1000 K.
Hot CO$_2$ is also important for industrial applications on Earth \citep{12EvFaCl.CO2} and studies
of combustion engines \citep{rs10}. These applications all require
information on CO$_2$ spectra at higher temperatures. It is this
problem that we address here.

%Ames improvements \citep{19HuScLe.CO2,17HuScFr.CO2}
%CDSD-200 update \citep{19TaPeGa.CO2}
%OCO-2 parameters \citep{16CoBoDu.CO2,17OyPaDr.CO2}

%Venus is 95\%\ CO2. CO2 opacity of Venus
%Industrial applications \citep{12EvFaCl.CO2}
%engine \cite{rs10}

%Comprehensive study of CO2 exoplanet abundances \cite{16HeLyxx.CO2}
%CO2 important measure of C/O ratio in hot Jupiters\citep{13MoKaVi}
%key marker in directly imaged planets \citep{16MoMaZa}
%lower mass planets atmospheres dominated by water and CO2 \citep{16MaHaTi}
%detection in exoplanets \citep{09SwTiVa.CO2,09SwVaTi.CO2,sdg10}.

The importance of CO$_2$ has led to very significant activity on
the construction of list of important rotation-vibration transition lines.
The HITRAN data base provides such lists for studies of the Earth's
atmosphere and other applications at or below 300 K. The
CO$_2$ line lists  where comprehensively updated in the 2016 release
of HITRAN \citep{jt691s} in part to provide higher accuracy data
to meet the demands of Earth observation satellites such as OCO-2
 \citep{16CoBoDu.CO2,17OyPaDr.CO2}. The 2016 update made extensive use
 of variational nuclear motion calculations \citep{jt625,jt667,jt678} of
 the type employed here.

 HITRAN is not designed for or suitable for high temperature applications
 which demand much more extensive line lists. The HITEMP data base \citep{jt480} is designed to address this issue. The original HITEMP
 used the direct numerical diagonalization calculations   of \citet{92WaRoxx.CO2}, which were an early example of  the use of large scale variational nuclear motion calculations to provide molecular line lists. The 2010 HITEMP update used the CDSD-1000 (carbon dioxide spectroscopic databank) line list \citep{02TaPeTe.CO2}. The CDSD-1000 line list is
 based on the use effective Hamiltonian fits to experimental data and
 was designed to be complete for temperatures up to 1000 K. CDSD-1000
 subsequently was replaced by CDSD-4000. The empirical CO$_2$ line list
 CDSD-4000 computed by \citet{11TaPexx.CO2} is designed for temperatures up to 5000~K, but has  limited wavenumber coverage. Furthermore, while usually good
 at reproducing known spectra, experience has shown that the effective
 Hamiltonian approach can struggle to capture all the unobserved hot bands resulting in underestimates of the opacity at higher temperatures \citep{jt780}.
 A compact version of CDSD-4000 has recently been made available  \citep{20VaLoSi}.

The NASA Ames group have produced a number of CO\2\ line lists \citep{13HuFrTa.CO2,14HuGaFr.CO2,17HuScFr.CO2,19HuScLe.CO2}
using highly accurate potential energy surfaces \citep{12HuScTa.CO2,17HuScFr.CO2}  and variational
nuclear motion calculations. Most relevant for this work is the
Ames-2016 CO\2\ line list of \citet{17HuScFr.CO2} which considers  wavenumbers  up to 15~000 \cm\ and $J$ up to 150 with upper state energies limited to $hc\times$24~000 \cm. Due to this fixed upper energy cut-off, the temperature coverage depends on the wavenumber range, from $T\sim 4000$~K at lower end up to $\sim 1500$~K at the higher end. The Ames-2016 line list was based on an accurate empirically-generated potential energy (PES) surface Ames-2 and their high-level \ai\ dipole moment surface (DMS) DMS-N2.

High accuracy room temperature line lists for 13 isotopologues of CO\2\ was computed by \citet{jt625,jt667,jt678}, using the Ames-2 PES \citep{17HuScFr.CO2} and UCL's highly accurate DMS \citep{jt613}. These line lists are now part of the HITRAN \citep{jt691s} and ExoMol \citep{jt528} databases. The room temperature properties of this line list have been
subject to a number experimental tests and the results have been found to be competitive in accuracy to state-of-the-art laboratory experiments \citep{jt700,18KaWaSu.CO2,18CeKaMo.CO2,20LaReFl.CO2}.

Here we present a new hot line list for the main isotopologue of CO\2\ (\COtwo) generated using UCL's \ai\ DMS \citep{jt613} and the empirical PES Ames-2 \citep{17HuScFr.CO2}  with the variational program TROVE \citep{TROVE}. Our line list is the most comprehensive (complete and accurate) data set for CO\2.  This work is performed as part of the
ExoMol project \citep{jt528} and the results form an important
addition to the ExoMol database \citep{jt631} which, as discussed
below, is currently being upgraded \citep{jt810}.

\section{TROVE specifications}

For this work we used a new implementation of the exact kinetic energy (EKE) operator for triatomics in TROVE \citep{20YuMexx} based on the bisector embedding for triatomic molecules \citep{83CaHaSu,jt96}.

The variational TROVE program \citep{TROVE} solves the ro-vibrational Schr\"{o}dinger equation using a multi-layer contraction scheme (see, for example, \citet{17YuYaOv.methods}). At step 1, the 1D primitive basis set functions $\phi_{v_1}(r_1)$, $\phi_{v_2}(r_2)$ (stretching) and $\phi_{v_3}(\rho)$ (bending) are obtained by numerically solving the corresponding Schr\"{o}dinger equations. Here $r_1$ and $r_2$ are two stretching valence coordinates and $\rho = 180^{\circ}-\alpha$ with $\alpha$ being the inter-bond valence angle. A 1D Hamiltonian operator for a given mode is constructed by setting all other degrees of freedom to the their equilibrium values. The two equivalent stretching equations are solved on a grid of 1000 points using the Numerov-Cooley approach \citep{24Numerov.method,61Cooley.method},  with the grid values of $r_i$ ranging from $r\rme-0.4$ to $r\rme+1.0$ \AA. The bending mode solutions are obtained on the basis of the associated Laguerre polynomials
as given by
\begin{equation}
\label{e:Laguerre}
\phi_{n,l}^{(l)}(\rho)  = C_{n,l} \, \rho^{l+1/2}\, L_{n}^{(l)}(a\rho^2)\, e^{-a \rho^2/2},
\end{equation}
normalized as
$$
\int_{0}^{\rho_{\rm max}} \phi_{n,l}^{(l)}(\rho)^2 \, d \rho = 1,
$$
%with
%$$
%C_{n,l} = \sqrt{\frac{2\sqrt{a}}{(n+l)!}} a^\frac{2 l +1}{4},
%$$
where $a$ is a structural parameter, $l \ge 0$, $\rho_{\rm max}$ was set to 170$^{\circ}$ and all primitive bending functions were mapped on a grid of 3000 points. The kinetic energy operator is constructed numerically as a formal expansion in terms of the inverse powers of the stretching coordinates $r_i$ ($i=1,2$): $1/r_i$ and $1/r_i^2$ around a non-rigid configuration \citep{70HoBuJo} defined by the $\rho_i$ points on the grid.
The singularities of the kinetic energy operator at $\rho=0^{\circ}$ ($\sim 1/\rho$ and $1/\rho^2$) are resolved analytically with the help of the
factors $\rho^{l+1/2}$ in the definition of the associated Laguerre basis set in Eq.~\eqref{e:Laguerre}.
The details of the model will be published elsewhere \citep{20YuMexx}.

At step 2 two reduced problems for the 2D stretching and 1D bending reduced Hamiltonians are solved variationally on the primitive basis set of $\ket{v_1,v_2} = \phi_{v_1}(r_1)\phi_{v_2}(r_2)$  and $\ket{v_3,l}  = \phi_{v_3}^{(l)}(\rho)$, respectively. The reduced Hamiltonians are
constructed by averaging the 3D vibrational ($J=0$) Hamiltonian over the ground state basis functions as follows:
\begin{align}
\label{e:H(1)}
\hat{H}_{\rm str}^{(1)}(r_1,r_2) &= \bra{0_3,l=0} \hat{H}^{\rm 3D} \ket{0_3,l=0},  \\
\label{e:H(2)}
\hat{H}_{\rm bnd}^{(2)}(\rho) &= \bra{0_1,0_2}  \hat{H}^{\rm 3D}  \ket{0_1,0_2},
\end{align}
where  $\ket{v_1,v_2}$ are  stretching and $\ket{v_3,l}$ are bending vibrational basis functions with $v_i=0$ and $l=0$. In the bending basis set, $l$ is treated as a parameter with the corresponding Hamiltonian matrices
$$
{H}_{i,j}^{(2),l} = \bra{i,l} \hat{H}_{\rm bnd}^{(2)}(\rho) \ket{j,l},
$$
block-diagonal in $l$ ($l=0,\ldots,l_{\rm max}$). The eigenfunctions of the reduced Hamiltonians in Eqs.~(\ref{e:H(1)},\ref{e:H(2)}), $\Phi^{(1)}_{i_1}(r_1,r_2)$ and $\Phi^{(2)}_{i_2,l}(\rho)$ are obtained variationally and then symmetrized using the automatic symmetry adaptation technique \citep{17YuYaOv.methods}.  A 3D vibrational basis set for the $J=0$ Hamiltonian for step 3 is then formed  as symmetry adapted products given by:
\begin{equation}
\label{e:contr:basis}
\Phi_{i_1,i_2,l}^{\Gamma_{\rm vib}} = \{\Phi^{(1)}_{i_1}(r_1,r_2) \, \Phi^{(2)}_{i_2,l}(\rho) \}^{\Gamma_{\rm vib}},
\end{equation}
where $\Gamma_{\rm vib}$ is the vibrational symmetry in the \Cv{2}(M) molecular symmetry group  \citep{98BuJe.method} used to classify the irreducible representations (irreps) of the ro-vibrational states of CO\2. \Cv{2}(M) comprises four irreps $A_1$, $A_2$, $B_1$ and $B_2$. The allowed vibrational symmetries are $A_1$ and $B_2$. The allowed ro-vibrational symmetries of \COtwo\ are $A_1$ and $A_2$ due to the restriction on the nuclear-spin-ro-vibrational functions imposed by the nuclear spin statistics (Pauli exclusion principle).

At step 3, the vibrational $(J=0)$ Hamiltonians are solved on the symmetry adapted vibrational basis in Eq.~\eqref{e:contr:basis}.
% for each value of $l$\footnote{In practice, for technical reasons, we combine $l$ and $v_3$ into a generalized quantum number $v_l = v_3*(l_{\rm max}+1)+l$ and solve one large matrix with block-diagonal in $l$. Here $l_{\rm max}$ is the maximal value assumed for $l$.}.
These eigenfunction $\Psi_{\lambda,l}^{(J=0)}$ are parameterized with $l$ and associated with a vibrational symmetry $A_1$ or $B_2$ and then used to build a ro-vibrational basis set ($J\ge 0$) as a symmetrized product:
\begin{equation}
\label{e:Psi-basis}
\Psi_{\lambda,K}^{(J,\Gamma)} = \{ \Psi_{\lambda,K}^{(J=0,\Gamma_{\rm vib})} \, \ket{J,K,\Gamma_{\rm rot}} \}^{\Gamma},
\end{equation}
where the rotational part $\ket{J,K,\Gamma_{\rm rot}}$ is a symmetrized combination of the rigid rotor functions \citep{17YuYaOv.methods} and  the rotational quantum number $K$ ($K\ge 0$) is constrained to the vibrational parameter $l$ ($K=l$).

An $E$ = $hc \times$36~000~\cm\ energy cut-off was used to contract the $J=0$ eigenfunctions. All energies and eigenfunctions up to $J=202$ were generated and used to produce the dipole lists for CO\2.

The size of the vibrational basis was controlled by a polyad-number condition:
$$
P  =  v_1+v_2 + v_3 \le  P_{\rm max} = 64,
$$
chosen based on the convergence tests with $v_1$ and $v_2 \le 56$.

\begin{figure}
\centering
\includegraphics[width=0.9\columnwidth]{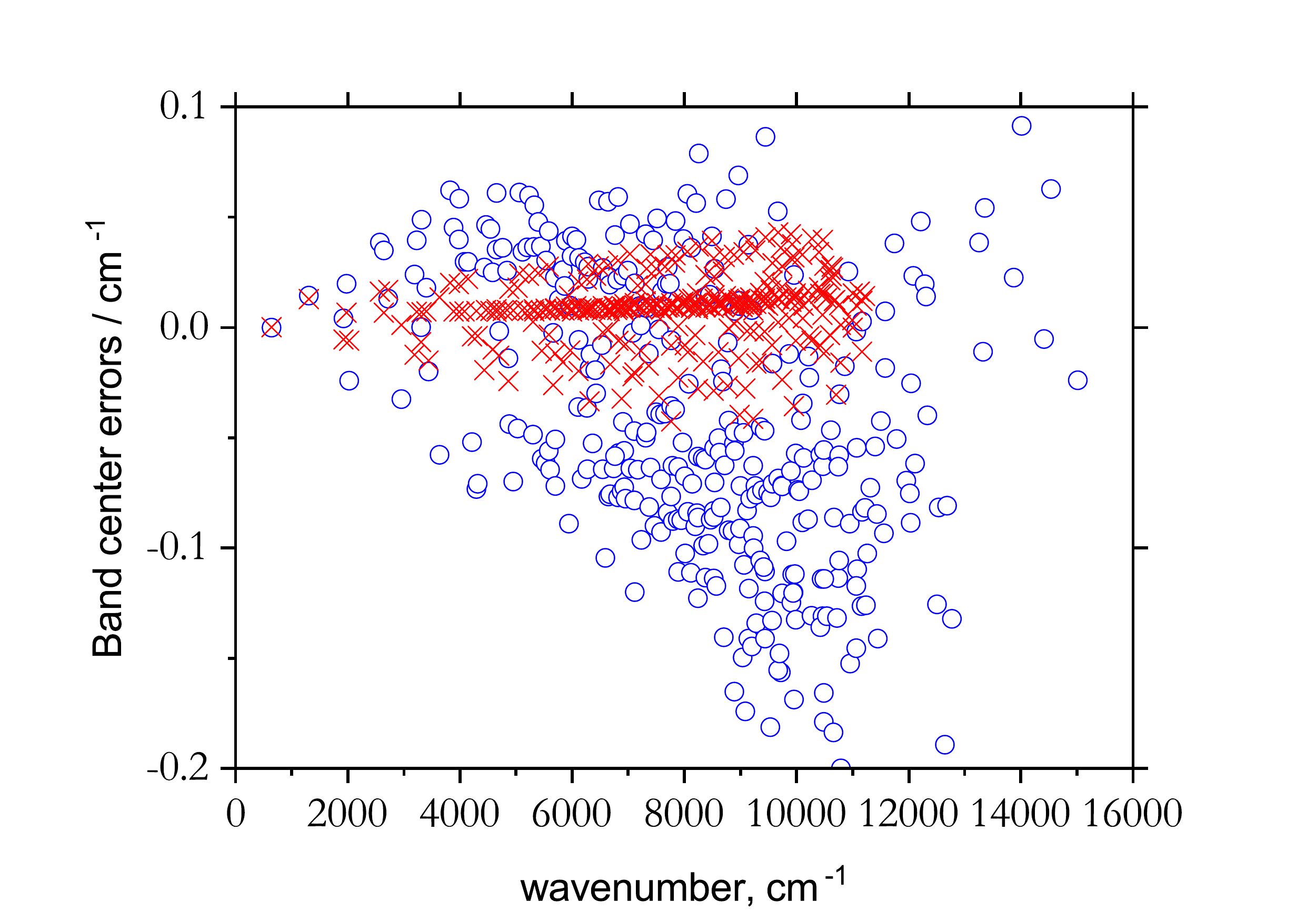}
\caption{Blue circles: Average ro-vibrational errors for 337 vibrational ($J=0$) states from HITRAN estimated as CO\2\ $ \tilde{E}^{\rm H}_\lambda-\tilde{E}^{\rm T}_\lambda$ (HITRAN-TROVE) term values for $J\le 40$, which were used as band centre corrections (see text). Red crosses: Average residuals for the ro-vibrational CO\2\ term values (for $J\le 40$) between TROVE (before the band centre correction) and calculations from \citet{jt625} using DVR3D and the same PES Ames-2016 for 294 vibrational states. }
\label{f:obs-calc}
\end{figure}

Some vibrational energies computed using the Ames-2 PES by \citet{17HuScFr.CO2} are shown in Table~\ref{t:bandcenters} and compared to the empirical \COtwo\ band centres (HITRAN's estimates, see below).
In order to improve the accuracy of the ro-vibrational energies, we have applied vibrational band centre corrections to the TROVE $J=0$ energies by shifting them to the HITRAN values, where available (see \citet{jt500}). This trick in combination with the $J=0$ contracted basis set allowed us to replace the diagonal vibrational matrix elements in the ro-vibrational Hamiltonian by the corresponding empirical band centres; for this reason the approach was named by \citet{jt500} the empirical basis set correction (EBSC). The band centre corrections were estimated as average residuals $\tilde{E}_i^{\rm T}-\tilde{E}_i^{\rm H}$ ($J=0,\ldots,40$) by matching the TROVE  $\tilde{E}_i^{\rm T}$  and HITRAN $\tilde{E}_i^{\rm H}$ ro-vibrational term values for $J\le 40$, wherever available, for each vibrational state present in HITRAN.  In total, 337 band centres\footnote{The band centres in question correspond to the fundamental or overtone bands and represent pure vibrational ($J=0$) term values. } ranging up to $\sim 15\,500$~\cm\ were corrected, with the total root-mean-square (rms) error of 0.06~\cm. This is illustrated in Table~\ref{t:bandcenters}, where 60 lowest term values before and after correction are shown together with the three alternative assignment cases, and in Fig.~\ref{f:obs-calc}, where the average ro-vibrational errors for 337 bands are plotted. The complete list of the band centres and their corrections is given as supplementary material to the paper.

\begin{table}
\caption{Examples of vibrational band centres of CO\2\ computed using Ames-2 PES with TROVE ($\tilde{E}^{\rm T}_\lambda $), average differences with the HITRAN band centres ($\Delta \tilde{E}_\lambda = \tilde{E}^{\rm H}_\lambda-\tilde{E}^{\rm T}_\lambda$),
the shifted band canters adopted for ro-vibrational calculations ($\tilde{E}_\lambda^{\rm T}+\Delta \tilde{E}_\lambda$) and three sets of quantum numbers (QN) used in this work. See the complete table in supplementary material. }
\tabcolsep=5pt
\label{t:bandcenters}
\begin{tabular}{rrrrrrrrrrrrrrrrr}
\hline\hline
& \multicolumn{4}{c}{Linear Molecule QN} & \multicolumn{5}{c}{HITRAN QN} &\multicolumn{3}{c}{TROVE QN}&\multicolumn{3}{c}{ Term values (\cm)}\\
\cline{2-5}
\cline{11-13}
$\lambda$  &  $n_1$& \multicolumn{1}{c}{$n_2^{\rm lin}$} &  $l_2$& $n_3$&  $m_1$ &  $m_2$&  $l_2$&  $m_3$ &  $r$ &  $v_1$ & $v_2$ &$v_3$  &    $\tilde{E}_\lambda^{\rm T}$ &$\Delta \tilde{E}_\lambda $  &$\tilde{E}_\lambda^{\rm T}+\Delta \tilde{E}_\lambda$ \\
\hline
     1 &    0 &    0 &    0 &    0 &    0 &    0 &    0 &    0 &    1 &     0 &    0 &    0 &        0.000 &     0.000 &          0.000   \\
     2 &    0 &    1 &    1 &    0 &    0 &    1 &    1 &    0 &    1 &     0 &    0 &    0 &      667.755 &     0.015 &        667.769   \\
     3 &    0 &    2 &    0 &    0 &    1 &    0 &    0 &    0 &    2 &     0 &    0 &    1 &     1285.404 &     0.004 &       1285.408   \\
     4 &    0 &    2 &    2 &    0 &    0 &    2 &    2 &    0 &    1 &     0 &    0 &    0 &     1336.673 &     0.020 &       1336.693   \\
     5 &    1 &    0 &    0 &    0 &    1 &    0 &    0 &    0 &    1 &     1 &    0 &    0 &     1388.209 &    -0.024 &       1388.185   \\
     6 &    0 &    3 &    1 &    0 &    1 &    1 &    1 &    0 &    2 &     0 &    0 &    1 &     1932.821 &     0.039 &       1932.860   \\
     7 &    0 &    3 &    3 &    0 &    0 &    3 &    3 &    0 &    1 &     0 &    0 &    0 &     2006.732 &     0.035 &       2006.766   \\
     8 &    1 &    1 &    1 &    0 &    1 &    1 &    1 &    0 &    1 &     1 &    0 &    0 &     2077.233 &     0.013 &       2077.246   \\
     9 &    0 &    0 &    0 &    1 &    0 &    0 &    0 &    1 &    1 &     0 &    1 &    0 &     2349.174 &    -0.032 &       2349.141   \\
    10 &    1 &    2 &    0 &    0 &    2 &    0 &    0 &    0 &    3 &     1 &    0 &    1 &     2548.343 &     0.024 &       2548.367   \\
    11 &    0 &    4 &    2 &    0 &    1 &    2 &    2 &    0 &    2 &     0 &    0 &    1 &     2586.546 &     0.039 &       2586.585   \\
    12 &    2 &    0 &    0 &    0 &    2 &    0 &    0 &    0 &    2 &     1 &    1 &    0 &     2671.144 &     0.000 &       2671.144   \\
    13 &    0 &    4 &    4 &    0 &    0 &    4 &    4 &    0 &    1 &     0 &    0 &    0 &     2677.936 &     0.049 &       2677.985   \\
    14 &    1 &    2 &    2 &    0 &    1 &    2 &    2 &    0 &    1 &     1 &    0 &    0 &     2762.269 &     0.018 &       2762.287   \\
    15 &    1 &    2 &    0 &    0 &    2 &    0 &    0 &    0 &    1 &     1 &    0 &    1 &     2797.157 &    -0.020 &       2797.137   \\
    16 &    0 &    1 &    1 &    1 &    0 &    1 &    1 &    1 &    1 &     0 &    1 &    0 &     3004.453 &    -0.058 &       3004.395   \\
    17 &    1 &    3 &    1 &    0 &    2 &    1 &    1 &    0 &    3 &     1 &    0 &    1 &     3181.792 &     0.062 &       3181.854   \\
    18 &    0 &    5 &    3 &    0 &    1 &    3 &    3 &    0 &    2 &     0 &    0 &    1 &     3244.101 &     0.045 &       3244.146   \\
    19 &    2 &    1 &    1 &    0 &    2 &    1 &    1 &    0 &    2 &     1 &    1 &    0 &     3339.706 &     0.040 &       3339.746   \\
    20 &    0 &    5 &    5 &    0 &    0 &    5 &    5 &    0 &    1 &     0 &    0 &    0 &     3350.284 &     0.058 &       3350.343   \\
    21 &    1 &    3 &    3 &    0 &    1 &    3 &    3 &    0 &    1 &     1 &    0 &    0 &     3445.706 &     0.030 &       3445.735   \\
    22 &    1 &    3 &    1 &    0 &    2 &    1 &    1 &    0 &    1 &     1 &    0 &    1 &     3501.033 &     0.030 &       3501.063   \\
    23 &    0 &    2 &    0 &    1 &    1 &    0 &    0 &    1 &    2 &     0 &    1 &    1 &     3612.891 &    -0.052 &       3612.839   \\
    24 &    0 &    2 &    2 &    1 &    0 &    2 &    2 &    1 &    1 &     0 &    1 &    0 &     3660.884 &    -0.073 &       3660.811   \\
    25 &    1 &    0 &    0 &    1 &    1 &    0 &    0 &    1 &    1 &     2 &    0 &    0 &     3714.852 &    -0.071 &       3714.781   \\
    26 &    1 &    4 &    0 &    0 &    3 &    0 &    0 &    0 &    4 &     1 &    0 &    2 &     3792.656 &     0.027 &       3792.683   \\
    27 &    1 &    4 &    2 &    0 &    2 &    2 &    2 &    0 &    3 &     1 &    0 &    1 &     3823.531 &     0.046 &       3823.577   \\
    28 &    0 &    6 &    4 &    0 &    1 &    4 &    4 &    0 &    2 &     0 &    0 &    1 &     3904.544 &     0.045 &       3904.589   \\
    29 &    3 &    0 &    0 &    0 &    3 &    0 &    0 &    0 &    3 &     1 &    2 &    0 &     3942.517 &     0.025 &       3942.542   \\
    30 &    2 &    2 &    2 &    0 &    2 &    2 &    2 &    0 &    2 &     1 &    1 &    0 &     4009.441 &     0.035 &       4009.476   \\
    31 &    0 &    6 &    6 &    0 &    0 &    6 &    6 &    0 &    1 &     0 &    0 &    0 &     4023.775 &     0.061 &       4023.836   \\
    32 &    3 &    0 &    0 &    0 &    3 &    0 &    0 &    0 &    2 &     1 &    2 &    0 &     4064.277 &    -0.002 &       4064.275   \\
    33 &    1 &    4 &    4 &    0 &    1 &    4 &    4 &    0 &    1 &     1 &    0 &    0 &     4128.500 &     0.036 &       4128.536   \\
    34 &    1 &    4 &    2 &    0 &    2 &    2 &    2 &    0 &    1 &     1 &    0 &    1 &     4198.897 &     0.026 &       4198.923   \\
    35 &    1 &    4 &    0 &    0 &    3 &    0 &    0 &    0 &    1 &     1 &    0 &    2 &     4225.111 &    -0.014 &       4225.097   \\
    36 &    0 &    3 &    1 &    1 &    1 &    1 &    1 &    1 &    2 &     0 &    1 &    1 &     4248.131 &    -0.044 &       4248.088   \\
    37 &    0 &    3 &    3 &    1 &    0 &    3 &    3 &    1 &    1 &     0 &    1 &    0 &     4318.455 &    -0.070 &       4318.386   \\
    38 &    1 &    1 &    1 &    1 &    1 &    1 &    1 &    1 &    1 &     2 &    0 &    0 &     4391.058 &    -0.046 &       4391.012   \\
    39 &    1 &    5 &    1 &    0 &    3 &    1 &    1 &    0 &    4 &     1 &    0 &    2 &     4416.480 &     0.061 &       4416.541   \\
    40 &    1 &    5 &    3 &    0 &    2 &    3 &    3 &    0 &    3 &     1 &    0 &    1 &     4470.630 &     0.034 &       4470.665   \\
    41 &    0 &    7 &    5 &    0 &    1 &    5 &    5 &    0 &    2 &     0 &    0 &    1 &     4567.381 &     0.036 &       4567.417   \\
    42 &    3 &    1 &    1 &    0 &    3 &    1 &    1 &    0 &    3 &     1 &    2 &    0 &     4591.448 &     0.060 &       4591.508   \\
    43 &    0 &    0 &    0 &    2 &    0 &    0 &    0 &    2 &    1 &     1 &    1 &    0 &     4673.371 &    -0.049 &       4673.322   \\
    44 &    2 &    3 &    3 &    0 &    2 &    3 &    3 &    0 &    2 &     1 &    1 &    0 &     4680.274 &     0.037 &       4680.311   \\
    45 &    0 &    7 &    7 &    0 &    0 &    7 &    7 &    0 &    1 &     0 &    0 &    0 &     4698.403 &     0.055 &       4698.458   \\
    46 &    3 &    1 &    1 &    0 &    3 &    1 &    1 &    0 &    2 &     1 &    2 &    0 &     4753.794 &     0.048 &       4753.841   \\
    47 &    1 &    5 &    5 &    0 &    1 &    5 &    5 &    0 &    1 &     1 &    0 &    0 &     4811.138 &     0.037 &       4811.175   \\
    48 &    1 &    2 &    0 &    1 &    2 &    0 &    0 &    1 &    3 &     2 &    0 &    1 &     4853.681 &    -0.060 &       4853.622   \\
    49 &    0 &    4 &    2 &    1 &    1 &    2 &    2 &    1 &    2 &     0 &    1 &    1 &     4889.592 &    -0.062 &       4889.531   \\
    50 &    1 &    5 &    3 &    0 &    2 &    3 &    3 &    0 &    1 &     1 &    0 &    1 &     4893.567 &     0.030 &       4893.597   \\
    51 &    1 &    5 &    1 &    0 &    3 &    1 &    1 &    0 &    1 &     1 &    0 &    2 &     4938.732 &     0.044 &       4938.775   \\
    52 &    0 &    4 &    4 &    1 &    0 &    4 &    4 &    1 &    1 &     0 &    1 &    0 &     4977.178 &    -0.065 &       4977.113   \\
    53 &    2 &    0 &    0 &    1 &    2 &    0 &    0 &    1 &    2 &     3 &    0 &    0 &     4977.889 &    -0.056 &       4977.833   \\
    54 &    1 &    6 &    0 &    0 &    4 &    0 &    0 &    0 &    5 &     1 &    0 &    3 &     5022.354 &    -0.002 &       5022.352   \\
    55 &    1 &    6 &    2 &    0 &    3 &    2 &    2 &    0 &    4 &     1 &    0 &    2 &     5048.803 &     0.023 &       5048.825   \\
    56 &    1 &    2 &    2 &    1 &    1 &    2 &    2 &    1 &    1 &     2 &    0 &    0 &     5063.368 &    -0.051 &       5063.317   \\
    57 &    1 &    2 &    0 &    1 &    2 &    0 &    0 &    1 &    1 &     2 &    0 &    1 &     5099.731 &    -0.072 &       5099.659   \\
    58 &    1 &    6 &    4 &    0 &    2 &    4 &    4 &    0 &    3 &     1 &    0 &    1 &     5121.776 &     0.013 &       5121.788   \\
    59 &    3 &    2 &    0 &    0 &    4 &    0 &    0 &    0 &    4 &     1 &    2 &    1 &     5197.228 &     0.026 &       5197.255   \\
    60 &    0 &    8 &    6 &    0 &    1 &    6 &    6 &    0 &    2 &     0 &    0 &    1 &     5232.314 &     0.019 &       5232.333   \\
\hline\hline
\end{tabular}
\end{table}

As an independent benchmark of the TROVE calculations, the initial TROVE energies ($J\le 40$) before the band centre corrections were compared to the theoretical CO\2\ energies computed by \citet{jt625} using the DVR3D program \citep{jt338} and the same PES as the Ames-2016 line list. Figure~\ref{f:obs-calc} shows averaged residuals for 294 bands matched to the CO\2\ energy term values from \citet{jt625} up to 10~500~\cm\ with the total rms error of 0.02~\cm. 

The ro-vibrational energies  were computed variationally using  the CO\2\ empirical PES Ames-2 by \citet{17HuScFr.CO2} for $J=0\ldots 230$ and used for the temperature  partition function of CO\2.  The transitional intensities (Einstein~A coefficients) were then computed using the UCL \ai\ DMS \citep{jt613} covering the wavenumber range from 0 to 20~000~\cm\   with the lower energy term value up to 16~000~\cm\ ($J\le 202$). To speed up the calculation of the  dipole moment matrix elements, a threshold of $10^{-12}$ to the eigen-coefficients was applied (see  \citet{jt500}). We have also applied a threshold of $10^{-8}$~Debye to the vibrational matrix elements of the dipole moment in order to reduce an accumulation error for higher overtones, see discussion by \citet{16MeMeSt,jt794}.

\section{Line list}

\subsection{Quantum numbers}

In TROVE calculations, the ro-vibrational states are uniquely identified by three numbers, the rotational angular momentum quantum number $J$, the total symmetry $\Gamma$ (Molecular symmetry group) and the eigen-state counting number $\lambda$ (in the order of increasing energies). Each state can be further assigned with approximate quantum numbers (QN) associated with the corresponding largest basis set contribution \citep{TROVE}. There are two main sets of approximate  QNs corresponding to the contractions steps 1 and 3.  The first set is connected to the primitive basis set excitation numbers $v_1, v_2, v_3$ and $l_2$. The second  set is associated with the vibrational counting number $\lambda$  from the ($J=0$)-contracted basis set Eq.~\eqref{e:Psi-basis}. Even though $v_1, v_2, v_3$ and $l_2$ are more physically intuitive than the counting number of the vibrational states $\lambda$, the latter is useful for correlating TROVE's states to the experimental (e.g. normal mode) QNs or indeed to any other scheme. Table~\ref{t:bandcenters} shows vibrational ($J=0$) term values of CO\2\ together with all assignment schemes either used in these work or relevant to the spectroscopy of CO\2: (i) TROVE primitive QNs $v_1, v_2, v_3$ and $l$, (ii) TROVE band centres counting number $\lambda$, (iii) HITRAN QNs $m_1$, $m_2$, $l_2$, $m_3$, $r$ adopted for CO\2\ and (iv) spectroscopic (normal mode) QNs $n_1$, $n_2^{\rm lin}$, $l_2$, $n_3$ used for  other  general linear triatomic molecules. Our preferred choice is (iv). According to this convention, $n_1$ and $n_3$ are two stretching quantum numbers associated with the symmetric and asymmetric modes; $n_2^{\rm lin}$ is the (symmetric) linear molecule bending quantum number; $l_2$ is the bending quantum number  satisfying  the standard conditions on the vibrational angular momentum of an isotropic 2D Harmonic oscillator \citep{98BuJe.method}
$$
l_2 = n_2^{\rm lin}, n_2^{\rm lin}-2, \ldots\, 1 (0).
$$

All ro-vibrational states are automatically assigned the QN schemes (i) and (ii), which were then automatically correlated to the general linear molecule QN (iv) using the following rules:
$$
l = l_2, \quad  v_1 + v_2 = n_1 + n_3,  \quad v_3 = n_2^{\rm lin},
$$
for a given stretching polyad $n_1+n_2$, where we assumed that the asymmetric quanta has higher energies than symmetric. For example, for $v_1+v_2=n_1+n_3=2$ the stretching QNs ($n_1,n_3$) and assigned to the vibrational term values according with the order of energies: $(2,0)$ 3339.702~\cm, $(1,1)$ 3714.853~\cm\ and $(0,2)$ 4673.392~\cm.
The linear molecule bending quantum number $n_2^{\rm lin}$ is given by
$$
n_2^{\rm lin} = 2 v_3 + l_2,
$$
where $v_3$ is the TROVE vibrational bending quantum number. The standard linear molecule QN scheme (iv) was favoured for example by \citet{53HeHe.CO2} and is also recommended here.

We could not perform similar automatic correlation to the HITRAN quantum labels $(m_1,m_2,l_2,m_3,r)$ for all the states in our line lists, only for states present in the CO\2\ HITRAN database. The HITRAN convention of quantum numbers (iii) for CO\2\ is more empirical. It is motivated by energy clusters formed by states in accidental Fermi resonance and their order within a cluster \citep{81RoYoxx.CO2}.  The quantum number $m_1$ is associated with Fermi resonance groups of (symmetric) states of different combinations of $l_2$ and $2 m_1 + m_2 = 2 n_1 + n_2^{\rm lin}$;
$r$ is the ranking index, with $r=1$ for the highest vibrational level of a Fermi resonance group and assuming  the values $1,2,\ldots m_1+1$ \citep{81RoYoxx.CO2}. HITRAN's version of the bending quantum number $m_2$ is  by definition equal  to $l_2$  (so-called AFGL notation) and thus redundant, while the stretching quantum number $m_3$ is the same as the linear molecule asymmetric quantum number $n_3$ from scheme (iv).  In order to simplify correlation with HITRAN and experimental literature, the scheme (iii) is retained, but only for the ro-vibrational states present in HITRAN.

%\red{sy: Even after consulting with \citet{81RoYoxx.CO2}, the HITRAN scheme is still confusing because it does not agree with it: according to  \citet{81RoYoxx.CO2}, $m_1, m_2, l_2, m_3$ should be standard linear molecule $n_1,n_2^{lin}, l_2,m_3$, but they are very different. For example the condition $m_2 = l_2$ does not make sense.}

\section{Line list}

The CO\2\  UCL-4000 line list contains 3~480~477 states and 2~557~551~923 transitions and covers the wavenumber range from 0 to 20~000~\cm\  (wavelengths, $\lambda>0.5$~\um) with the lower energy up to 16~000~\cm\ and $J\le 202$ with a threshold on the Einstein coefficients of $10^{-14}$~s$^{-1}$.
The line list consists of two files \citep{jt631}, called States and Transitions, which in the case of UCL-4000 are summarised in Tables~\ref{t:states} and \ref{t:trans}. The first 4 columns of the States file have the compulsory structure for all molecules: State ID, Energy term value (\cm), the total degeneracy and the total angular momentum.
According to the new ExoMol-2020 format \citep{jt810}, the 5th column is also compulsory representing the uncertainty estimate of the corresponding term value (\cm), which is followed by the lifetime, Land\'{e} $g$-factor (if provided) and  molecular specific quantum numbers, including rigorous (symmetry, parity) and non-rigorous (vibrational, rotational, etc). The States file covers all states up to $J=230$ (3~526~057 states). The Transitions part consists of  three columns with the upper State ID, lower State ID and the Einstein $A$ coefficient, see Table~\ref{t:trans}. For convenience, the Transition part is split into 20 files each covering 1000~\cm.

\begin{table}
{\tt
\caption{Extracts from the final states file for UCL-4000. }
\label{t:states}
\tabcolsep=5pt
\begin{tabular}{rrrrrccrrrrrrrrrrrrrcrc}
\hline\hline
$i$ & \multicolumn{1}{c}{$\tilde{E}$} & $g_{\rm tot}$  & $J$ & \multicolumn{1}{c}{unc.} & $\Gamma$ &$e/f$ &  $n_1$ &$n_2^{\rm lin}$ & $l_2$ &$n_3$& $C_i$ &$m_1$ &$m_2$ & $m_3$ &$m_4$& $r$&$v_1^{\rm T}$ &$v_2^{\rm T}$  &$v_3^{\rm T}$ \\
%& $\Gamma_{\rm vib}$ & $K$ & $\Gamma_{\rm rot}$\\
\hline
           1&     0.000000&      1&       0&       0.0005& A1&  e &   0 &  0 &  0 &  0 &  1.00 &     0 &  0&   0 &  0 &  1&      0&   0&   0\\
           2&  1285.408200&      1&       0&       0.0005& A1&  e &   0 &  2 &  0 &  0 &  1.00 &     1 &  0&   0 &  0 &  2&      0&   0&   1\\
           3&  1388.184200&      1&       0&       0.0050& A1&  e &   1 &  0 &  0 &  0 &  1.00 &     1 &  0&   0 &  0 &  1&      1&   0&   0\\
           4&  2548.366700&      1&       0&       0.0005& A1&  e &   1 &  2 &  0 &  0 &  1.00 &     2 &  0&   0 &  0 &  3&      1&   0&   1\\
           5&  2671.142957&      1&       0&       0.0050& A1&  e &   2 &  0 &  0 &  0 &  1.00 &     2 &  0&   0 &  0 &  2&      1&   1&   0\\
           6&  2797.136000&      1&       0&       0.0050& A1&  e &   1 &  2 &  0 &  0 &  1.00 &     2 &  0&   0 &  0 &  1&      1&   0&   1\\
           7&  3792.681898&      1&       0&       0.0050& A1&  e &   1 &  4 &  0 &  0 &  1.00 &     3 &  0&   0 &  0 &  4&      1&   0&   2\\
           8&  3942.541358&      1&       0&       0.0050& A1&  e &   3 &  0 &  0 &  0 &  1.00 &     3 &  0&   0 &  0 &  3&      1&   2&   0\\
           9&  4064.274256&      1&       0&       0.0050& A1&  e &   3 &  0 &  0 &  0 &  1.00 &     3 &  0&   0 &  0 &  2&      1&   2&   0\\
          10&  4225.096148&      1&       0&       0.0050& A1&  e &   1 &  4 &  0 &  0 &  1.00 &     3 &  0&   0 &  0 &  1&      1&   0&   2\\
          11&  4673.325200&      1&       0&       0.0005& A1&  e &   0 &  0 &  0 &  2 &  1.00 &     0 &  0&   0 &  2 &  1&      1&   1&   0\\
          12&  5022.349428&      1&       0&       0.0050& A1&  e &   1 &  6 &  0 &  0 &  1.00 &     4 &  0&   0 &  0 &  5&      1&   0&   3\\
          13&  5197.252900&      1&       0&       0.0050& A1&  e &   3 &  2 &  0 &  0 &  1.00 &     4 &  0&   0 &  0 &  4&      1&   2&   1\\
          14&  5329.645446&      1&       0&       0.0050& A1&  e &   4 &  0 &  0 &  0 &  1.00 &     4 &  0&   0 &  0 &  3&      2&   2&   0\\
          15&  5475.553054&      1&       0&       0.0005& A1&  e &   3 &  2 &  0 &  0 &  1.00 &     4 &  0&   0 &  0 &  2&      1&   2&   1\\
          16&  5667.644584&      1&       0&       0.0050& A1&  e &   2 &  4 &  0 &  0 &  1.00 &     4 &  0&   0 &  0 &  1&      1&   1&   2\\
          17&  5915.212302&      1&       0&       0.0005& A1&  e &   0 &  2 &  0 &  2 &  1.00 &     1 &  0&   0 &  2 &  2&      1&   1&   1\\
          18&  6016.690121&      1&       0&       0.0005& A1&  e &   1 &  0 &  0 &  2 &  1.00 &     1 &  0&   0 &  2 &  1&      0&   3&   0\\
          19&  6240.044061&      1&       0&         0.08& A1&  e &   2 &  6 &  0 &  0 &  1.00 &    -1 & -1&  -1 & -1 & -1&      1&   1&   3\\
          20&  6435.507278&      1&       0&         0.06& A1&  e &   4 &  2 &  0 &  0 &  1.00 &    -1 & -1&  -1 & -1 & -1&      2&   2&   1\\
          21&  6588.323819&      1&       0&         0.05& A1&  e &   5 &  0 &  0 &  0 &  1.00 &    -1 & -1&  -1 & -1 & -1&      3&   2&   0\\
\hline\hline
\end{tabular}
}
\mbox{}\\

{\flushleft
$i$:   State counting number.     \\
$\tilde{E}$: State energy in \cm. \\
$g_{\rm tot}$: Total state degeneracy.\\
$J$: Total angular momentum.            \\
unc.: Uncertainty \cm: The empirical (4 decimal places) and estimated (2 decimal places) values.     \\
$\Gamma$:   Total symmetry index in \Cv{2}(M)\\
$e/f$: Kronig rotationless parity \\
$n_1$: Normal mode stretching symmetry ($A_1$) quantum number. \\
$n_2^{\rm lin}$ : Normal mode linear molecule bending ($A_1$) quantum number. \\
$l_2$: Normal mode vibrational angular momentum quantum number. \\
$n_3$: Normal mode stretching asymmetric ($B1_1$) quantum number. \\
$C_i$: Coefficient with the largest contribution to the $(J=0)$ contracted set; $C_i\equiv =1$ for $J=0$. \\
$m_1$:   CO\2\ symmetric vibrational Fermi-resonance group quantum number (-1 stands for non available).\\
$m_2$:   CO\2\ vibrational bending quantum number (similar but not exactly identical to $n_2^{\rm lin}$).\\
$l_2$:   CO\2\ vibrational bending quantum number (similar but not exactly identical to standard $l_2$).\\
$m_3$:   CO\2\ asymmetric vibrational stretching quantum number, the same as $n_3$.\\
$r$:   CO\2\ additional ranking quantum number identifying states within a Fermi-resonance group.\\
$v_1^{\rm T}$:   TROVE stretching vibrational quantum number.\\
$v_2^{\rm T}$:   TROVE stretching vibrational quantum number.\\
$v_3^{\rm T}$:   TROVE bending vibrational quantum number.\\
%$\Gamma_{\rm vib}$:   Vibrational symmetry index in \Cv{2}(M). \\
%$K$:       Projection of $J$ on molecular symmetry axis ($K=L$).\\
%$\Gamma_{\rm rot}$:   Rotational symmetry index in \Cv{2}(M).
}
\end{table}

\begin{table}
\caption{Extract from the transitions file for the UCL-4000 line list.  }
{\tt
\begin{tabular}{rrr}
\hline\hline
$f$ & $i$ & $A_{fi}$\\
\hline
     1176508  &    1137722& 2.8564E-02 \\
     1861958  &    1849078& 1.6327E-03 \\
      631907  &     665295& 8.1922E-12 \\
     1344897  &    1331267& 4.2334E-07 \\
      983465  &     944281& 1.4013E-08 \\
      183042  &     170520& 1.6345E-02 \\
     2695389  &    2668685& 5.9366E-07 \\
      811518  &     822542& 1.4353E-01 \\
      406949  &     369902& 2.3774E-02 \\
\hline\hline
\end{tabular}
}
\label{t:trans}
\mbox{}\\
{$f$}: Upper state counting number.  \\
{$i$}: Lower state counting number. \\
$A_{fi}$: Einstein-A coefficient in s$^{-1}$.\\
\end{table}

The uncertainty of the CO\2\ energies were estimated based on the two main criteria. For the states matched to and replaced by the HITRAN energies, the uncertainties $\sigma$ (in cm$^{-1}$) are taken as the HITRAN errors of the corresponding line positions  as specified by the HITRAN error codes \citep{jt350}.  For all other states the following conservative estimate is used:
$$
\sigma = 0.01\, n_1 + 0.01\, n_2^{\rm lin} + 0.01\, n_3.
$$

\begin{figure}
\centering
\includegraphics[width=0.45\columnwidth]{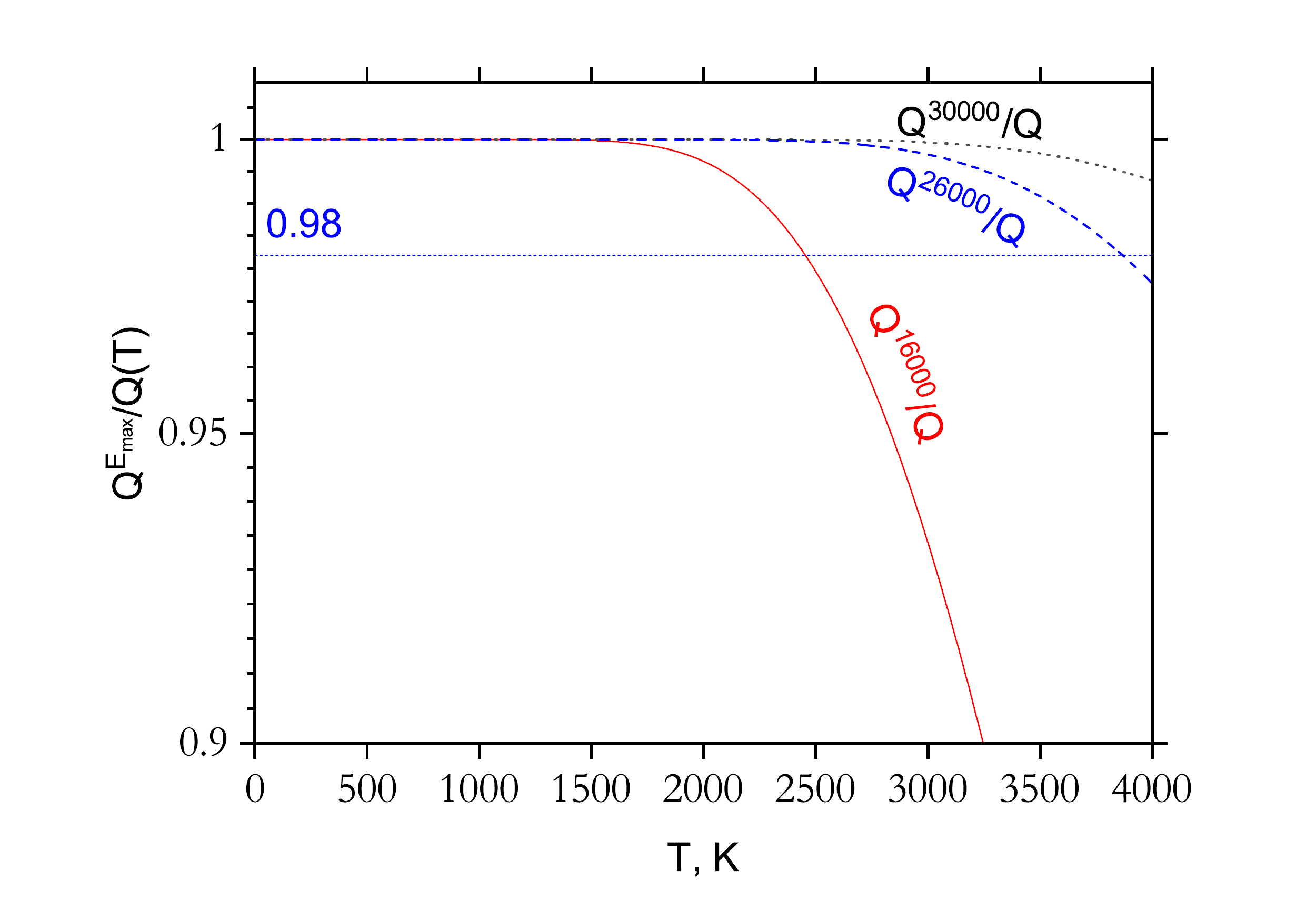}
\includegraphics[width=0.45\columnwidth]{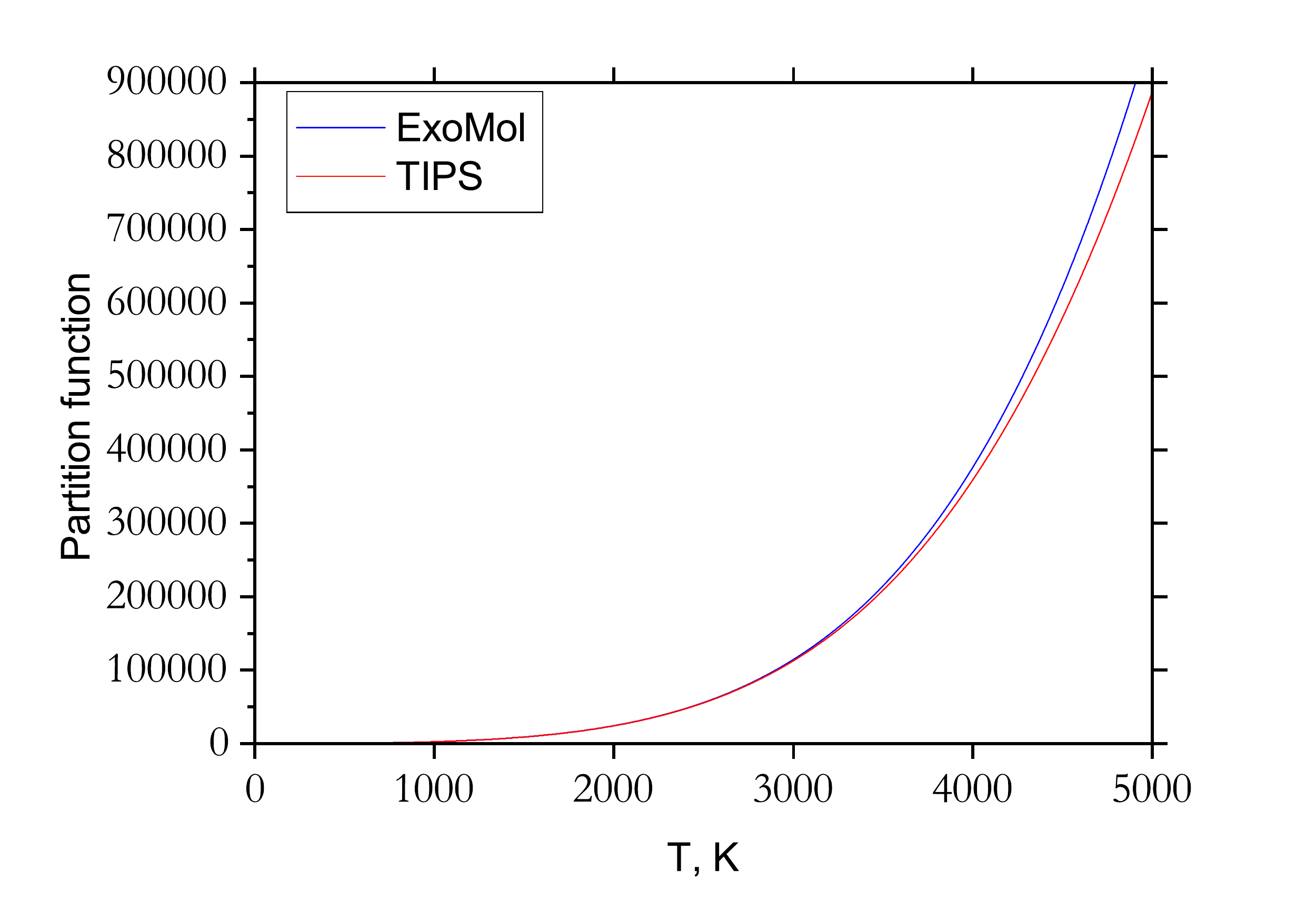}
\caption{Partition functions of CO\2. Left display: Ratio $Q^{E_{\rm max}}(T)/Q(T)$ of the CO\2\ partition functions at with energy cutoffs $hc\cdot$16~000~\cm,  $hc\cdot$26000~\cm\ and $hc\cdot$30000~\cm, as a function of temperature. $Q(T)$ is chosen as $Q^{36000~{\rm cm}^{-1}}$.  The line $0.98$ shows that the recommended temperature varies with the wavenumber region, from $T=$ 2500~K ($>$ 0~\cm) to $T=$ 4000~K ($>$ 10~000~\cm). Right display:  ExoMol values were computed using all UCL-4000 energies for $J \le 230$; TIPS values were obtained using  the TIPS app provided in \citet{TIPS2017}.}
\label{f:pf}
\end{figure}

With the  lower state energy threshold of  16~000~\cm\ our line list should be valid for temperatures up to at least 2500~K. Figure~\ref{f:pf} (left) illustrates the effect of the lack of the population for the states higher than 16~000~\cm\ with the help of the CO\2\ partition function $Q(T)$. In this figure we show a ratio  of  incomplete $Q^{E_{\rm max}}(T)$ (using only states below $E_{\rm max}$) over  `complete' $Q^{36000}$(T). This ratio has the difference of 2~\% for $Q^{16000}$ at $T=$ 2500~K and  $Q^{26000}$ at $T=$ 4000~K, which are the estimates for maximal temperature of the line list for wavenumber regions 0--20~000~\cm\ and 10~000--20~000~\cm, respectively. In Figure~\ref{f:pf} (right) we compare the partition functions of CO\2\ computed using our line list with  the Total Internal Partition Sums (TIPS) values for CO\2\ \citep{TIPS2017}. The energies in our line list cover higher rotational excitations ($J_{\rm max}$ = 230) compared to TIPS, which was based on the threshold of $J_{\rm max} =  150$. The energy thresholds in both cases are comparable, $hc \cdot$36~000~\cm\ vs $hc \cdot$30~383~\cm\ (TIPS). The threshold value of $J_{\rm max} =  150$ corresponds the energy term value of $\sim$ 10~000~\cm and thus leads the underestimate of the partition function of high $T$, while with the threshold of $J_{\rm max}$ = 230 all states below 20~000~\cm\ are included.

\begin{figure}
\centering
\includegraphics[width=0.9\columnwidth]{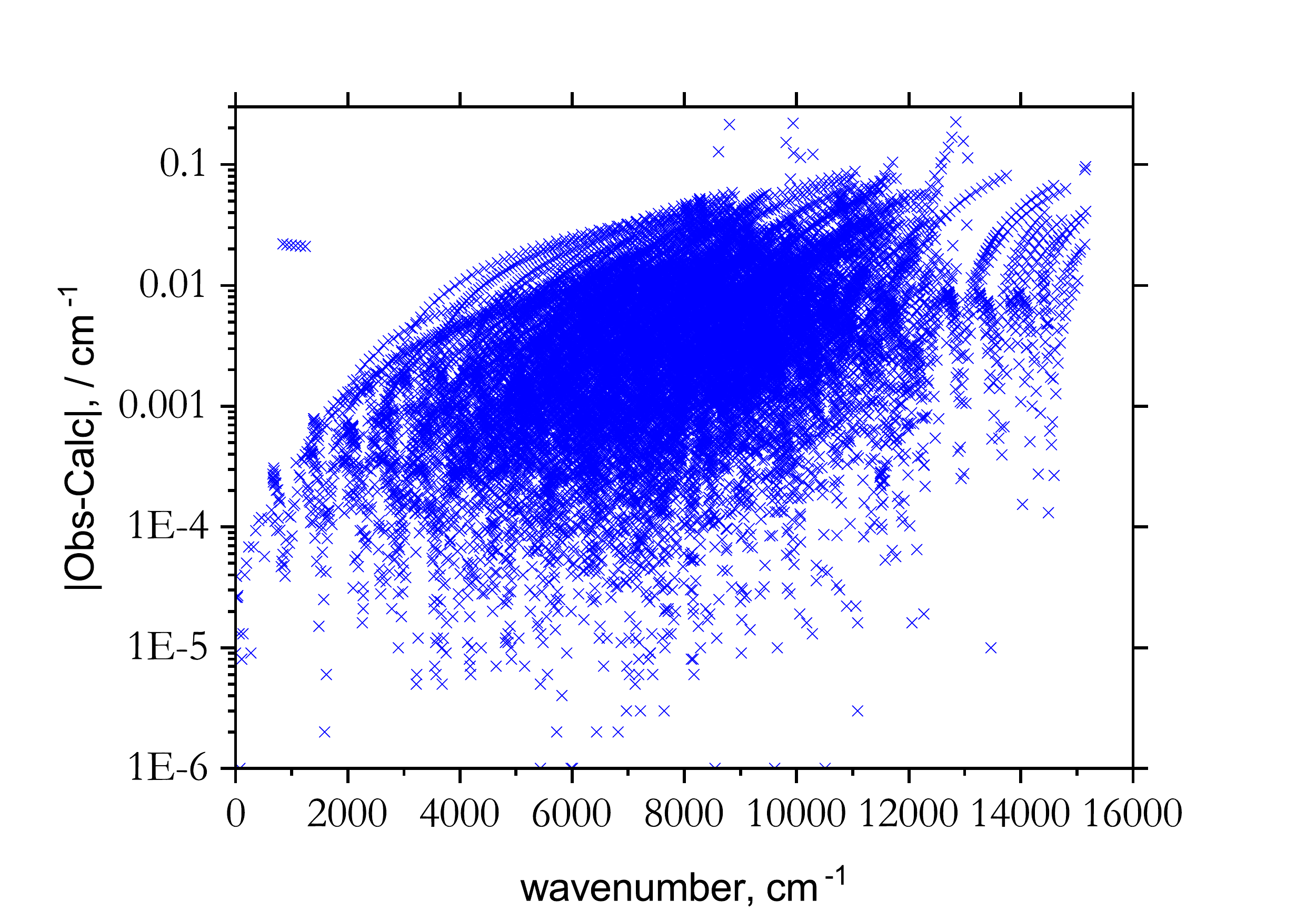}
\caption{Ro-vibrational Obs.-Calc. errors $|\tilde{E}_i^{\rm H}-\tilde{E}_i^{\rm T}|$ (HITRAN-TROVE)  in \cm\ for all HITRAN's 18~392 empirical values as a function of the energy term values, covering $J\le 129$ giving an rms error of 0.016~\cm.}
\label{f:rv:obs-calc}
\end{figure}

In order to improve the calculated line positions,  CO\2\ energies from HITRAN were used to replace the TROVE energies where available, taking advantage of the two-part structure of the UCL-4000 line list consisting of a States file and Transition files \citep{jt631}. A HITRAN energy list consisting of 18~392 empirical values from 337 vibrational states covering $J$ values up to 129 was generated by collecting all lower and upper state energies from the \COtwo\ HITRAN transitions. Comparison of the calculated TROVE term energies with these 18~668 HITRAN values gives an rms error of 0.016~\cm\ and is illustrated in Fig.~\ref{f:rv:obs-calc} in a log-scale.

\begin{figure}
\centering
\includegraphics[width=0.9\columnwidth]{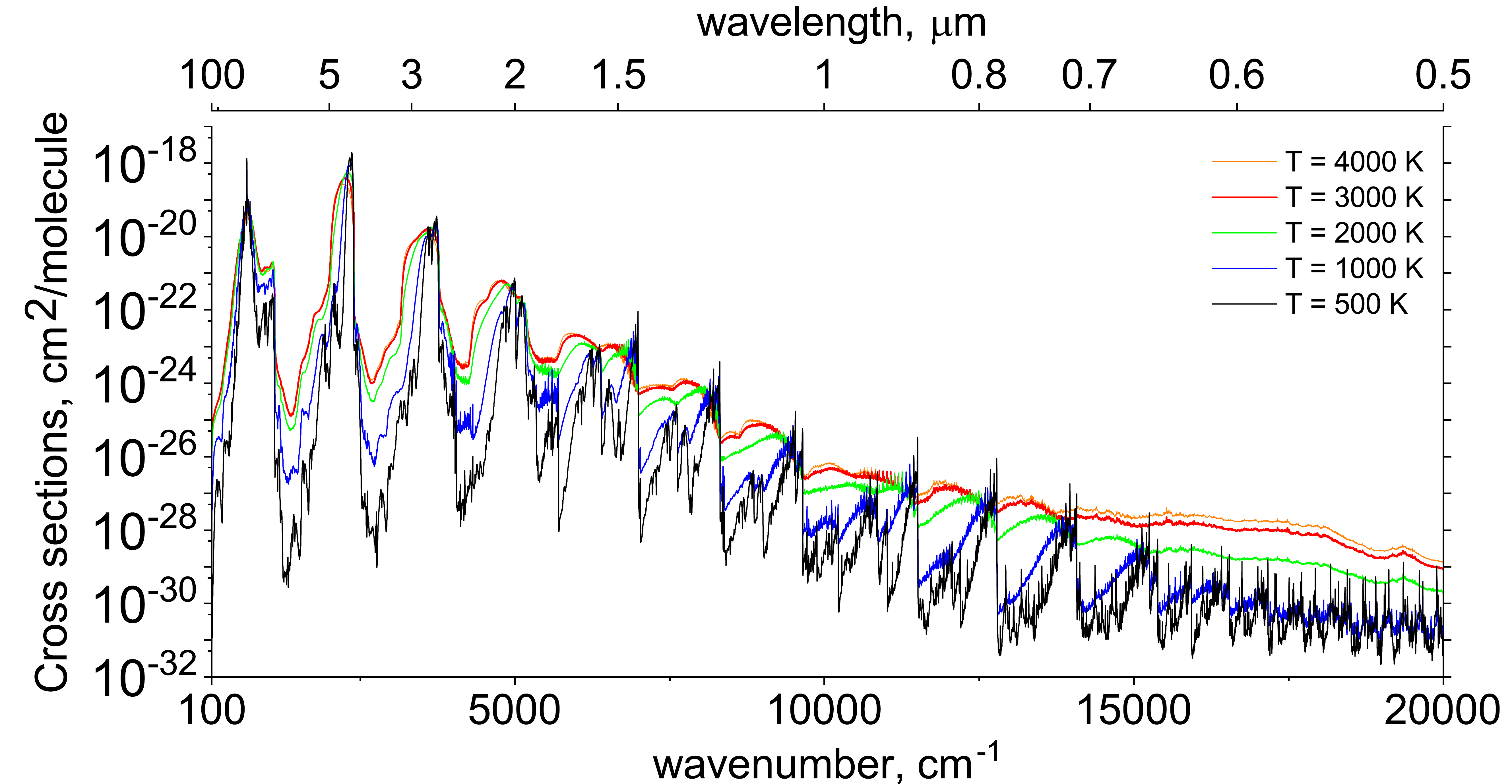}
\caption{Temperature dependence of the CO\2\ spectrum using the Gaussian line profile of HWHM=1~\cm. The spectrum becomes systematically flatter with increasing temperature.}
\label{fig:Temp}
\end{figure}

The overview of the CO\2\ absorption spectrum and its temperature dependence are illustrated in Fig.~\ref{fig:Temp}. CO\2\ has a prominent band at 4.3 \um, commonly used for atmospheric and astrophysical retrievals. For example, it was used as one of the photometric bands for the Spitzer Space Telescope \citep{Spitzer}.

\begin{figure}
\centering
\includegraphics[width=0.9\columnwidth]{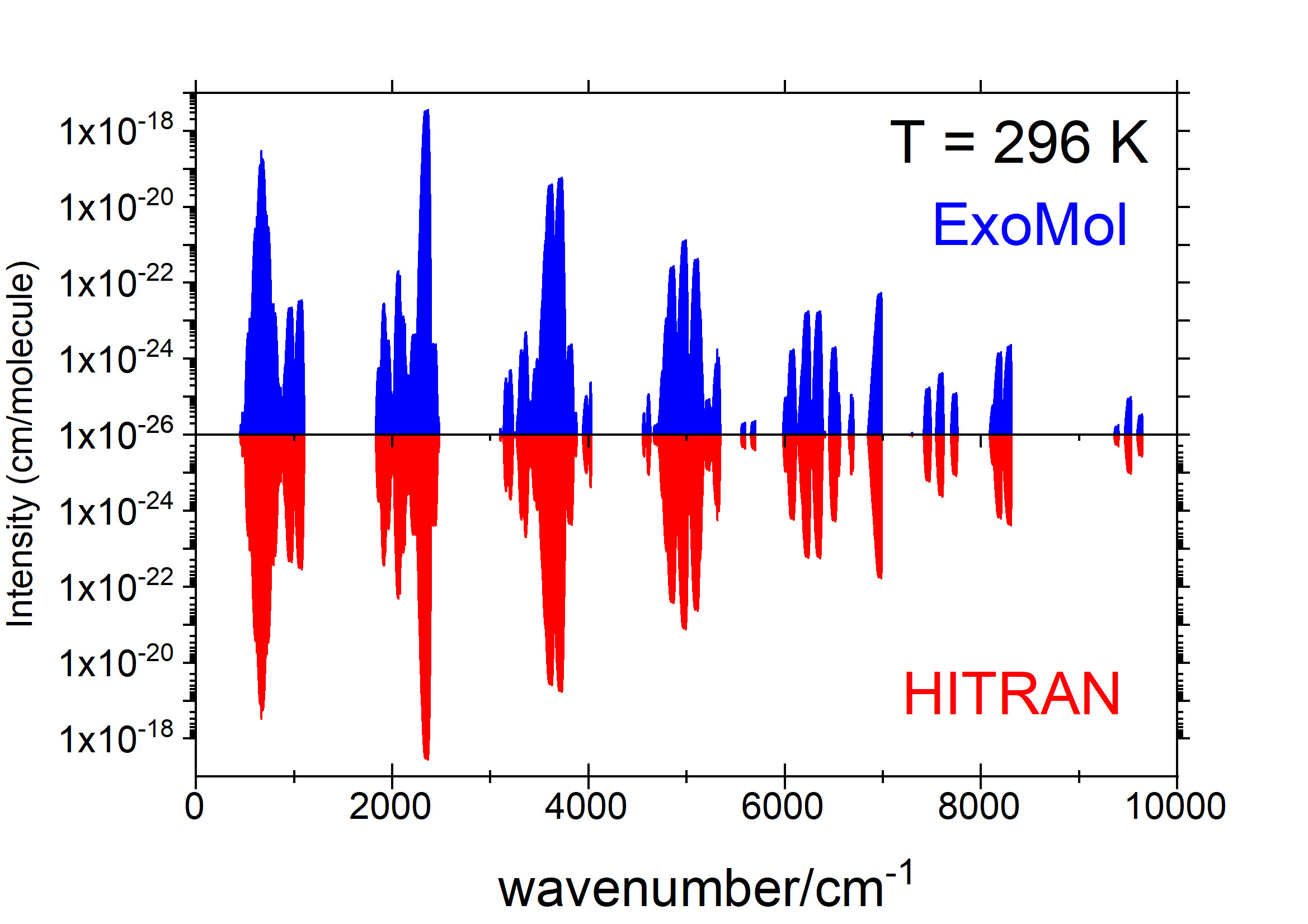}
\caption{CO\2\  stick spectrum comparing using UCL-4000 to HITRAN at $T=296$~K. }
\label{f:HITRAN:log}
\end{figure}

\begin{figure}
\centering
\includegraphics[width=0.45\columnwidth]{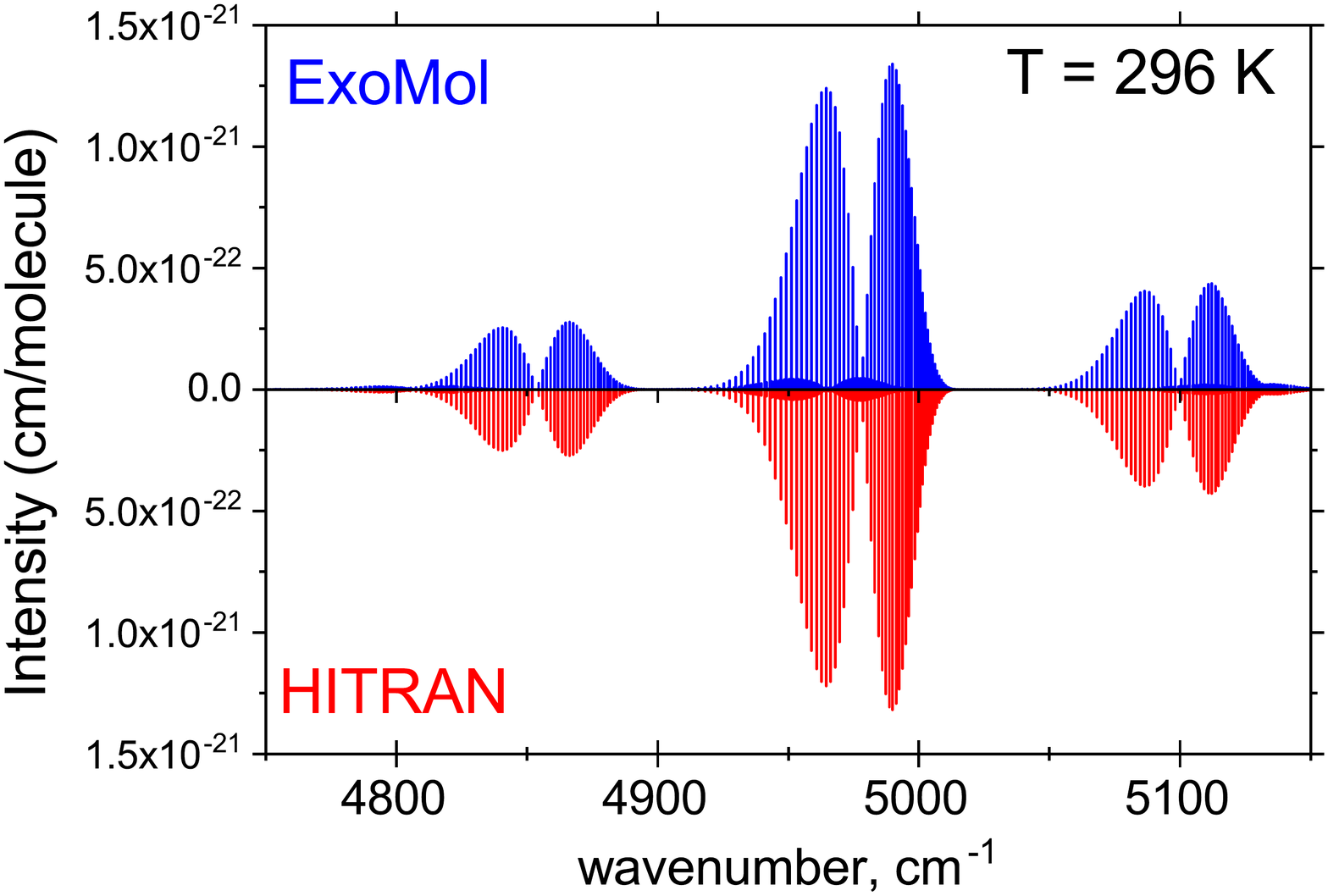}
\includegraphics[width=0.45\columnwidth]{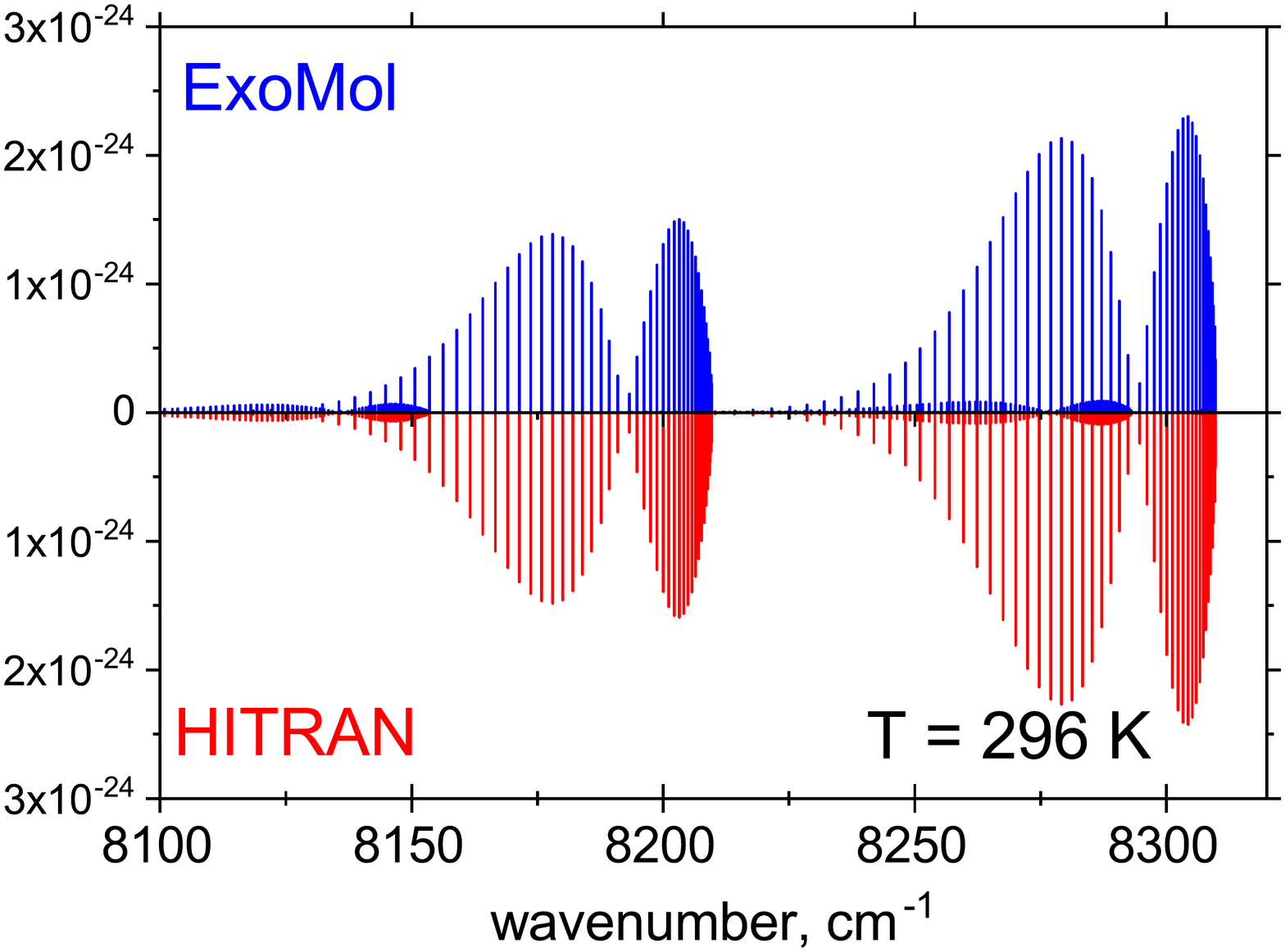}
\caption{Absorption stick spectrum of CO\2\ computed using UCL-4000 and compared to HITRAN at $T=296$~K for two spectroscopic regions.}
\label{f:HITRAN-1}
\end{figure}

Our line list is designed to almost perfectly agree with the HITRAN line positions, which was achieved by replacing the TROVE energies with the energies collected from the HITRAN data set for CO\2. Figures~\ref{f:HITRAN:log} and \ref{f:HITRAN-1} offer some comparisons with HITRAN.
In turn, the accuracy of the line intensities agree well with the HITRAN values as guaranteed by the quality of the UCL \ai\ DMS used. The mean ratio of our intensities to HITRAN is 1.0029 with the standard error of 0.00029 for  171143 HITRAN lines we could establish a correlation to.

\begin{figure}
\centering
\includegraphics[width=0.9\columnwidth]{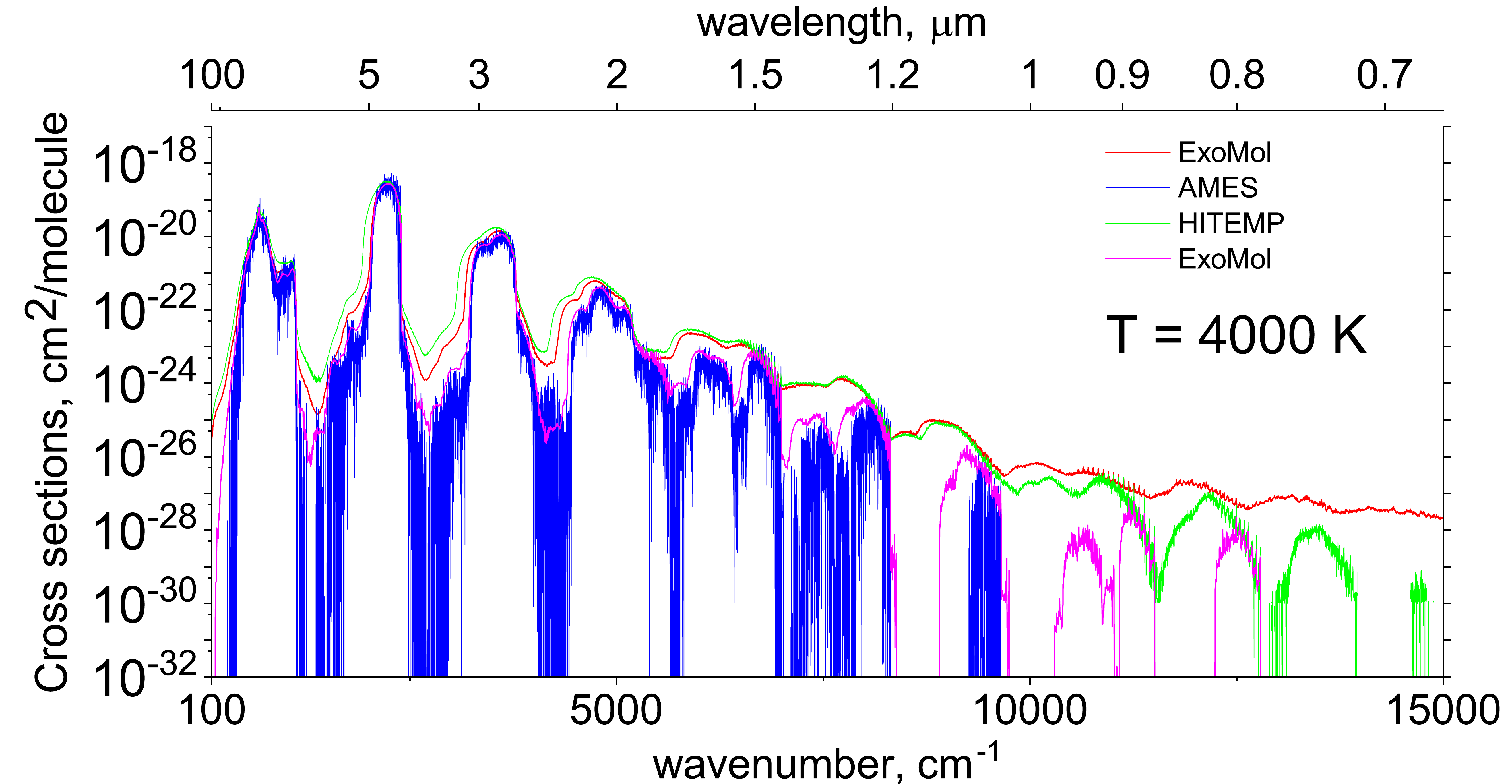}
\caption{Comparing UCL-4000, CDSD-4000 \citep{11TaPexx.CO2}, Ames-2016 \citep{17HuScFr.CO2} and 2010 HITEMP \citep{jt480} spectra of CO\2\ at $T=4000$~K. A Gaussian line profile of HWHM=1~\cm\ was used in each case.}
\label{f:CDSD:AMES}
\end{figure}

Figure~\ref{f:CDSD:AMES} compares the performance of the four main  line lists for hot CO\2, UCL-4000 (this work), CDSD-4000 (effective Hamiltonian) \citep{11TaPexx.CO2}, Ames-2 (variational) \citep{17HuScFr.CO2} and 2010 HITEMP (empirical)  \citep{jt480}  for the wavenumber range from 0 to 15~000~\cm. In general the UCL-4000 line list gives the highest opacity at $T=4000$~K which is due to its being more complete,  while both CDSD-4000 and 2010 HITEMP omit too many hot bands to be able to compete with the other two line lists at high $T$. The Ames-2016 line list also lacks a significant portion of the opacity at short wavelengths at this temperature as expected from the low energy threshold used \citep{17HuScFr.CO2}.

%Figure~\ref{f:H-E:H-H} compares the accuracy of the UCL-4000 line list to that of 2010 HITEMP as a difference from the room temperature HITRAN spectrum as the reference.

\begin{figure}
\centering
\includegraphics[width=0.45\columnwidth]{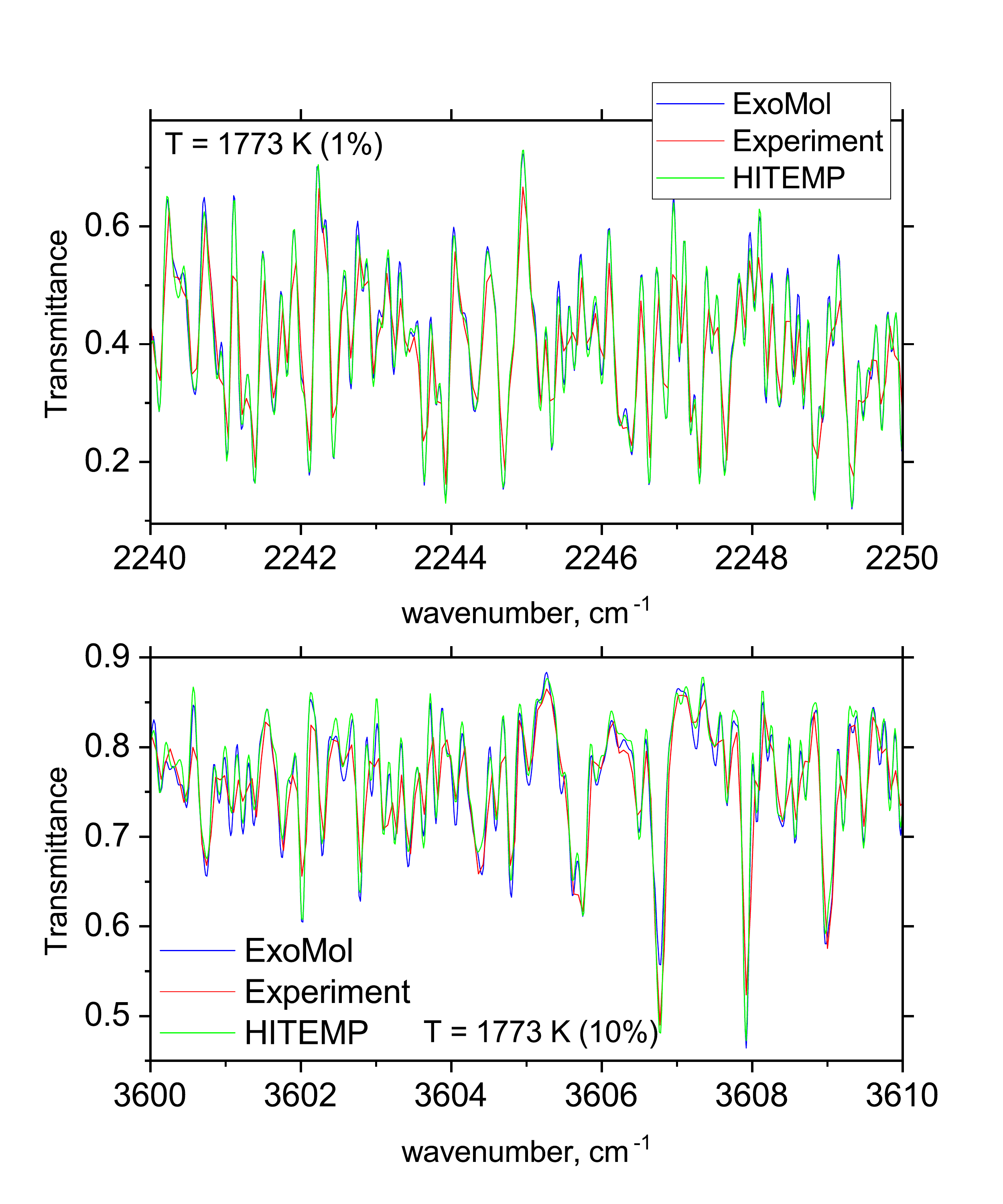}
\includegraphics[width=0.45\columnwidth]{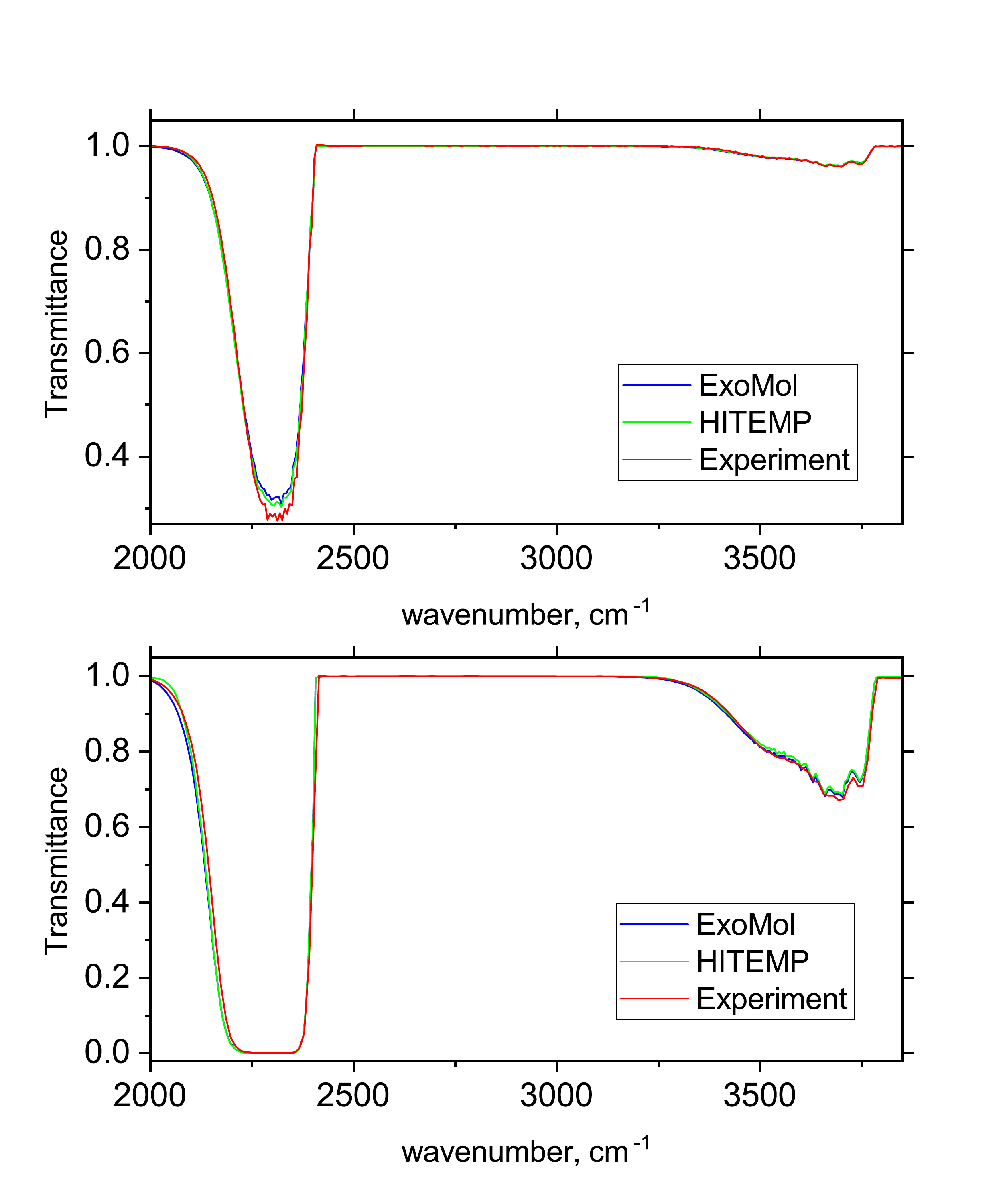}
\caption{Comparing HITEMP and ExoMol to experimental transmittance \citep{12EvFaCl.CO2} at $T=1773$~K and 1 bar using a Voigt profile. The experiments were performed with 1\%\ (upper) and 10\%\ (lower) CO$_2$ in N$_2$ buffer gas. HITEMP line width parameters were used for HITEMP, while for UCL-4000 we used an approximate scheme a0 from the ExoMol diet \citep{jt684} based on the HITRAN values. The left display shows two selected windows at  high resolution (0.0125~\cm), while the right display is with broader overview and lower resolution (6~\cm). }
\label{f:Fateev}
\end{figure}

Figure~\ref{f:Fateev} compares  the UCL-4000 spectrum of CO\2\ at $T=1773$~K  with the experiment by \citet{12EvFaCl.CO2}, who  recorded transmittance of CO\2\ at the normal pressure. The displays on the left show small windows from the  two strongest CO\2\ bands, 2.7~\um\ and 4.3~\um\ at higher resolution, while the displays on the right gives an overview of the whole region covering these bands at lower resolution. As a reference, spectra computed with the 2010 HITEMP line list are also shown. At this temperature,  UCL-4000 performs very similarity to 2010 HITEMP and shows excellent agreement with the experiment.

\section{Conclusion}

A new hot line list for the main isotopologue of CO\2\ (\COtwo) is presented, which is the most comprehensive (complete and accurate) data set for carbon dioxide to date. The line list is an important addition to the ExoMol
database which now contains line lists for all the major constituents
of hot Jupiter and mini-Neptune exoplanets. Line lists are still
being added to address the problem of hot super Earth exoplanets,
or lava planets, the composition of whose atmospheres are currently
not well constrained.

The line lists can be downloaded from the CDS (\url{http://cdsweb.u-strasbg.fr/}) or from ExoMol (\url{www.exomol.com} databases.

\section*{Acknowledgments}

This work was supported by the STFC Projects No. ST/M001334/1 and ST/R000476/1. The authors acknowledge the use of the UCL Legion High Performance Computing Facility (Legion@UCL) and associated support services in the completion of this work, along with the Cambridge Service for Data Driven Discovery (CSD3), part of which is operated by the University of Cambridge Research Computing on behalf of the STFC DiRAC HPC Facility (www.dirac.ac.uk). The DiRAC component of CSD3 was funded by BEIS capital funding via STFC capital grants ST/P002307/1 and ST/R002452/1 and STFC operations grant ST/R00689X/1. DiRAC is part of the National e-Infrastructure. We thank Alexander Fateev for the help with the experimental CO\2\ cross sections. SY acknowledges support from DESY.

\section*{Data availability statement}

Full data is made available. The line lists can be downloaded from the CDS (\url{http://cdsweb.u-strasbg.fr/}) or from ExoMol (\url{www.exomol.com} databases. 

The complete list of the band centres and their corrections is given as supplementary material to the paper

%\begin{figure}
%\centering
%\includegraphics[width=0.9\columnwidth]{CO2_AMES_TROVE-eps-converted-to.pdf}
%\caption{Comparing with AMES. $T = 4000$~K.}
%\label{fig:AMES}
%\end{figure}

%\begin{figure}
%\centering
%\includegraphics[width=0.9\columnwidth]{CO2_HITRAN-ExoMol-HITEMP-eps-converted-to.pdf}
%\caption{Comparing ExoMol and HITEMP to HITRAN at $T=296$~K. A Gaussian line profile of HWHM=0.1~\cm was used.}
%\label{f:H-E:H-H}
%\end{figure}

\label{lastpage}

\bibliographystyle{mn2e}
%\bibliography{journals_astro,jtj,methods,abinitio,programs,CO2,Books,linelists,partition,exogen,exoplanets,missions,sy}

\begin{thebibliography}{54}
\expandafter\ifx\csname natexlab\endcsname\relax\def\natexlab#1{#1}\fi

\bibitem[{Barton {et~al}\mbox{.}(2017)Barton, Hill, Czurylo, Li, Hyslop,
  Yurchenko, \& Tennyson}]{jt684}
Barton E.~J., Hill C., Czurylo M., Li H.-Y., Hyslop A., Yurchenko S.~N.,
  Tennyson J., 2017, J. Quant. Spectrosc. Radiat. Transf., 203, 490

\bibitem[{Baylis-Aguirre {et~al}\mbox{.}(2020)Baylis-Aguirre, Creech-Eakman, \&
  G\"{u}th}]{20BaCrGu.CO2}
Baylis-Aguirre D.~K., Creech-Eakman M.~J., G\"{u}th T., 2020, MNRAS, 493, 807

\bibitem[{Bunker \& Jensen(1998)}]{98BuJe.method}
Bunker P.~R., Jensen P., 1998, Molecular Symmetry and Spectroscopy, 2nd edn.
  NRC Research Press, Ottawa

\bibitem[{Carter {et~al}\mbox{.}(1983)Carter, Handy, \& Sutcliffe}]{83CaHaSu}
Carter S., Handy N., Sutcliffe B., 1983, Mol. Phys., 49, 745

\bibitem[{Chubb {et~al}\mbox{.}(2020)Chubb, Tennyson, \& Yurchenko}]{jt780}
Chubb K.~L., Tennyson J., Yurchenko S.~N., 2020, MNRAS, 493, 1531

\bibitem[{Connor {et~al}\mbox{.}(2016)Connor, B\"{o}sch, McDuffie, Taylor, Fu,
  Frankenberg, O'Dell, Payne, Gunson, Pollock, Hobbs, Oyafuso, \&
  Jiang}]{16CoBoDu.CO2}
Connor B. {et~al.}, 2016, Atmos. Meas. Tech., 9, 5227

\bibitem[{Cooley(1961)}]{61Cooley.method}
Cooley J.~W., 1961, Math. Comp., 15, 363

\bibitem[{Evseev {et~al}\mbox{.}(2012)Evseev, Fateev, \&
  Clausen}]{12EvFaCl.CO2}
Evseev V., Fateev A., Clausen S., 2012, J. Quant. Spectrosc. Radiat. Transf.,
  113, 2222

\bibitem[{Gamache {et~al}\mbox{.}(2017)Gamache, Roller, Lopes, Gordon, Rothman,
  Polyansky, Zobov, Kyuberis, Tennyson, Yurchenko, Cs{\'{a}}sz{\'{a}}r,
  Furtenbacher, Huang, Schwenke, Lee, Drouin, Tashkun, Perevalov, \&
  Kochanov}]{TIPS2017}
Gamache R.~R. {et~al.}, 2017, J. Quant. Spectrosc. Radiat. Transf., 203, 70

\bibitem[{Gordon \& {et al.}(2017)}]{jt691s}
Gordon I.~E., {et al.}, 2017, J. Quant. Spectrosc. Radiat. Transf., 203, 3

\bibitem[{Heng \& Lyons(2016)}]{16HeLyxx.CO2}
Heng K., Lyons J.~R., 2016, ApJ, 817, 149

\bibitem[{Herzberg \& Herzberg(1953)}]{53HeHe.CO2}
Herzberg G., Herzberg L., 1953, J. Opt. Soc. Am., 43, 1037

\bibitem[{Hougen {et~al}\mbox{.}(1970)Hougen, Bunker, \& Johns}]{70HoBuJo}
Hougen J.~T., Bunker P.~R., Johns J. W.~C., 1970, J. Mol. Spectrosc., 34, 136

\bibitem[{Huang {et~al}\mbox{.}({2013})Huang, Freedman, Tashkun, Schwenke, \&
  Lee}]{13HuFrTa.CO2}
Huang X., Freedman R.~S., Tashkun S.~A., Schwenke D.~W., Lee T.~J., {2013}, J.
  Quant. Spectrosc. Radiat. Transf., {130}, 134

\bibitem[{Huang {et~al}\mbox{.}({2014})Huang, Gamache, Freedman, Schwenke, \&
  Lee}]{14HuGaFr.CO2}
Huang X., Gamache R.~R., Freedman R.~S., Schwenke D.~W., Lee T.~J., {2014}, J.
  Quant. Spectrosc. Radiat. Transf., {147}, 134

\bibitem[{Huang {et~al}\mbox{.}(2017)Huang, Schwenke, Freedman, \&
  Lee}]{17HuScFr.CO2}
Huang X., Schwenke D.~W., Freedman R.~S., Lee T.~J., 2017, J. Quant. Spectrosc.
  Radiat. Transf., 203, 224

\bibitem[{Huang {et~al}\mbox{.}({2019})Huang, Schwenke, \& Lee}]{19HuScLe.CO2}
Huang X., Schwenke D.~W., Lee T.~J., {2019}, J. Quant. Spectrosc. Radiat.
  Transf., {230}, 222

\bibitem[{Huang {et~al}\mbox{.}(2012)Huang, Schwenke, Tashkun, \&
  Lee}]{12HuScTa.CO2}
Huang X., Schwenke D.~W., Tashkun S.~A., Lee T.~J., 2012, J. Chem. Phys., 136,
  124311

\bibitem[{Kang {et~al}\mbox{.}(2018)Kang, Wang, Liu, Sun, Zhou, Liu, \&
  Hu}]{18KaWaSu.CO2}
Kang P., Wang J., Liu G.-L., Sun Y.~R., Zhou Z.-Y., Liu A.-W., Hu S.-M., 2018,
  J. Quant. Spectrosc. Radiat. Transf., 207, 1

\bibitem[{Long {et~al}\mbox{.}(2020)Long, Reed, Fleisher, Mendonca, Roche, \&
  Hodges}]{20LaReFl.CO2}
Long D., Reed Z., Fleisher A., Mendonca J., Roche S., Hodges J., 2020, Geophys.
  Res. Lett., 47, e2019GL086344

\bibitem[{Massol {et~al}\mbox{.}({2016})Massol, Hamano, Tian, Ikoma, Abe,
  Chassefiere, Davaille, Genda, Guedel, Hori, Leblanc, Marcq, Sarda,
  Shematovich, Stoekl, \& Lammer}]{16MaHaTi}
Massol H. {et~al.}, {2016}, Space Sci. Rev., {205}, 153

\bibitem[{Medvedev {et~al}\mbox{.}(2016)Medvedev, Meshkov, Stolyarov, Ushakov,
  \& Gordon}]{16MeMeSt}
Medvedev E.~S., Meshkov V.~V., Stolyarov A.~V., Ushakov V.~G., Gordon I.~E.,
  2016, J. Mol. Spectrosc., 330, 36

\bibitem[{Medvedev {et~al}\mbox{.}(2020)Medvedev, Ushakov, Conway, Upadhyay,
  Gordon, \& Tennyson}]{jt794}
Medvedev E.~S., Ushakov V.~G., Conway E.~K., Upadhyay A., Gordon I.~E.,
  Tennyson J., 2020, J. Quant. Spectrosc. Radiat. Transf., 252, 107084

\bibitem[{Moses {et~al}\mbox{.}({2013})Moses, Madhusudhan, Visscher, \&
  Freedman}]{13MoKaVi}
Moses J.~I., Madhusudhan N., Visscher C., Freedman R.~S., {2013}, ApJ, {763},
  25

\bibitem[{Noumerov(1924)}]{24Numerov.method}
Noumerov B.~V., 1924, MNRAS, 84, 592

\bibitem[{Odintsova {et~al}\mbox{.}(2017)Odintsova, Fasci, Moretti, Zak,
  Polyansky, Tennyson, Gianfrani, \& Castrillo}]{jt700}
Odintsova T., Fasci E., Moretti L., Zak E.~J., Polyansky O.~L., Tennyson J.,
  Gianfrani L., Castrillo A., 2017, J. Chem. Phys., 146, 244309

\bibitem[{Oyafuso {et~al}\mbox{.}(2017)Oyafuso, Payne, Drouin, Devi, Benner,
  Sung, Yu, Gordon, Kochanov, Tan, Crisp, Mlawer, \& Guillaume}]{17OyPaDr.CO2}
Oyafuso F. {et~al.}, 2017, J. Quant. Spectrosc. Radiat. Transf., 203, 213

\bibitem[{Polyansky {et~al}\mbox{.}(2015)Polyansky, Bielska, Ghysels, Lodi,
  Zobov, Hodges, \& Tennyson}]{jt613}
Polyansky O.~L., Bielska K., Ghysels M., Lodi L., Zobov N.~F., Hodges J.~T.,
  Tennyson J., 2015, Phys. Rev. Lett., 114, 243001

\bibitem[{Rein \& Sanders(2010)}]{rs10}
Rein K.~D., Sanders S.~T., 2010, Appl. Optics, 49, 4728

\bibitem[{Rothman {et~al}\mbox{.}(2010)Rothman, Gordon, Barber, Dothe, Gamache,
  Goldman, Perevalov, Tashkun, \& Tennyson}]{jt480}
Rothman L.~S. {et~al.}, 2010, J. Quant. Spectrosc. Radiat. Transf., 111, 2139

\bibitem[{Rothman {et~al}\mbox{.}(2005)Rothman, Jacquemart, Barbe, Benner,
  Birk, Brown, Carleer, Chackerian, Chance, Coudert, Dana, Devi, Flaud,
  Gamache, Goldman, Hartmann, Jucks, Maki, Mandin, Massie, Orphal, Perrin,
  Rinsland, Smith, Tennyson, Tolchenov, Toth, Vander~Auwera, Varanasi, \&
  Wagner}]{jt350}
Rothman L.~S. {et~al.}, 2005, J. Quant. Spectrosc. Radiat. Transf., 96, 139

\bibitem[{Rothman \& Young(1981)}]{81RoYoxx.CO2}
Rothman L.~S., Young L.~D., 1981, J. Quant. Spectrosc. Radiat. Transf., 25, 505

\bibitem[{Snels {et~al}\mbox{.}({2014})Snels, Stefani, Grassi, Piccioni, \&
  Adriani}]{14SnStGr.CO2}
Snels M., Stefani S., Grassi D., Piccioni G., Adriani A., {2014}, Planet Space
  Sci., {103}, 347

\bibitem[{Sutcliffe \& Tennyson(1991)}]{jt96}
Sutcliffe B.~T., Tennyson J., 1991, Int. J. Quantum Chem., 39, 183

\bibitem[{Swain {et~al}\mbox{.}(2010)Swain, Deroo, Griffith, Tinetti, Thatte,
  Vasisht, Chen, Bouwman, Crossfield, Angerhausen, Afonso, \& Henning}]{sdg10}
Swain M.~R. {et~al.}, 2010, Nature, 463, 637

\bibitem[{Swain {et~al}\mbox{.}({2009}{\natexlab{a}})Swain, Tinetti, Vasisht,
  Deroo, Griffith, Bouwman, Chen, Yung, Burrows, Brown, Matthews, Rowe,
  Kuschnig, \& Angerhausen}]{09SwTiVa.CO2}
Swain M.~R. {et~al.}, {2009}{\natexlab{a}}, ApJ, {704}, 1616

\bibitem[{Swain {et~al}\mbox{.}({2009}{\natexlab{b}})Swain, Vasisht, Tinetti,
  Bouwman, Chen, Yung, Deming, \& Deroo}]{09SwVaTi.CO2}
Swain M.~R., Vasisht G., Tinetti G., Bouwman J., Chen P., Yung Y., Deming D.,
  Deroo P., {2009}{\natexlab{b}}, ApJL, {690}, L114

\bibitem[{Tashkun \& Perevalov(2011)}]{11TaPexx.CO2}
Tashkun S.~A., Perevalov V.~I., 2011, J. Quant. Spectrosc. Radiat. Transf.,
  112, 1403

\bibitem[{Tashkun {et~al}\mbox{.}({2003})Tashkun, Perevalov, Teffo, Bykov, \&
  Lavrentieva}]{02TaPeTe.CO2}
Tashkun S.~A., Perevalov V.~I., Teffo J.~L., Bykov A.~D., Lavrentieva N.~N.,
  {2003}, J. Quant. Spectrosc. Radiat. Transf., {82}, 165

\bibitem[{Tennyson {et~al}\mbox{.}(2004)Tennyson, Kostin, Barletta, Harris,
  Polyansky, Ramanlal, \& Zobov}]{jt338}
Tennyson J., Kostin M.~A., Barletta P., Harris G.~J., Polyansky O.~L., Ramanlal
  J., Zobov N.~F., 2004, Comput. Phys. Commun., 163, 85

\bibitem[{Tennyson \& Yurchenko(2012)}]{jt528}
Tennyson J., Yurchenko S.~N., 2012, MNRAS, 425, 21

\bibitem[{Tennyson {et~al}\mbox{.}(2020)Tennyson, Yurchenko, abd V.~H.
  J.~Clark, Chubb, Conway, Dewan, Gorman, Hill, Lynas-Gray, Mellor, McKemmish,
  Owens, Polyansky, Semenov, Somogyi, Tinetti, Upadhyay, Waldmann, Wang,
  Wright, \& Yurchenko}]{jt810}
Tennyson J. {et~al.}, 2020, J. Quant. Spectrosc. Radiat. Transf.

\bibitem[{Tennyson {et~al}\mbox{.}(2016)Tennyson, Yurchenko, Al-Refaie, Barton,
  Chubb, Coles, Diamantopoulou, Gorman, Hill, Lam, Lodi, McKemmish, Na, Owens,
  Polyansky, Rivlin, Sousa-Silva, Underwood, Yachmenev, \& Zak}]{jt631}
Tennyson J. {et~al.}, 2016, J. Mol. Spectrosc., 327, 73

\bibitem[{Vargas {et~al}\mbox{.}(2020)Vargas, Lopez, \& da~Silva}]{20VaLoSi}
Vargas J., Lopez B., da~Silva M.~L., 2020, J. Quant. Spectrosc. Radiat.
  Transf., 245, 106848

\bibitem[{\v{C}erm\'{a}k {et~al}\mbox{.}(2018)\v{C}erm\'{a}k, Karlovets,
  Mondelain, Kassi, Perevalov, \& Campargue}]{18CeKaMo.CO2}
\v{C}erm\'{a}k P., Karlovets E.~V., Mondelain D., Kassi S., Perevalov V.~I.,
  Campargue A., 2018, J. Quant. Spectrosc. Radiat. Transf., 207, 95

\bibitem[{Wattson \& Rothman(1992)}]{92WaRoxx.CO2}
Wattson R.~B., Rothman L.~S., 1992, J. Quant. Spectrosc. Radiat. Transf., 48,
  763

\bibitem[{Werner {et~al}\mbox{.}(2004)Werner, Roellig, Low, Rieke, Rieke,
  Hoffmann, Young, Houck, Brandl, Fazio, Hora, Gehrz, Helou, Soifer, Stauffer,
  Keene, Eisenhardt, Gallagher, Gautier, Irace, Lawrence, Simmons, Cleve, Jura,
  Wright, \& Cruikshank}]{Spitzer}
Werner M.~W. {et~al.}, 2004, ApJS, 154, 1

\bibitem[{Yurchenko {et~al}\mbox{.}(2011)Yurchenko, Barber, \&
  Tennyson}]{jt500}
Yurchenko S.~N., Barber R.~J., Tennyson J., 2011, MNRAS, 413, 1828

\bibitem[{Yurchenko \& Mellor(2020)}]{20YuMexx}
Yurchenko S.~N., Mellor T.~M., 2020, J. Chem. Phys., submitted

\bibitem[{Yurchenko {et~al}\mbox{.}(2007)Yurchenko, Thiel, \& Jensen}]{TROVE}
Yurchenko S.~N., Thiel W., Jensen P., 2007, J. Mol. Spectrosc., 245, 126

\bibitem[{Yurchenko {et~al}\mbox{.}(2017)Yurchenko, Yachmenev, \&
  Ovsyannikov}]{17YuYaOv.methods}
Yurchenko S.~N., Yachmenev A., Ovsyannikov R.~I., 2017, J. Chem. Theory
  Comput., 13, 4368

\bibitem[{Zak {et~al}\mbox{.}(2016)Zak, Tennyson, Polyansky, Lodi, Tashkun, \&
  Perevalov}]{jt625}
Zak E.~J., Tennyson J., Polyansky O.~L., Lodi L., Tashkun S.~A., Perevalov
  V.~I., 2016, J. Quant. Spectrosc. Radiat. Transf., 177, 31

\bibitem[{Zak {et~al}\mbox{.}(2017{\natexlab{a}})Zak, Tennyson, Polyansky,
  Lodi, Zobov, Tashkun, \& Perevalov}]{jt678}
Zak E.~J., Tennyson J., Polyansky O.~L., Lodi L., Zobov N.~F., Tashkun S.~A.,
  Perevalov V.~I., 2017{\natexlab{a}}, J. Quant. Spectrosc. Radiat. Transf.,
  203, 265

\bibitem[{Zak {et~al}\mbox{.}(2017{\natexlab{b}})Zak, Tennyson, Polyansky,
  Lodi, Zobov, Tashkun, \& Perevalov}]{jt667}
Zak E.~J., Tennyson J., Polyansky O.~L., Lodi L., Zobov N.~F., Tashkun S.~A.,
  Perevalov V.~I., 2017{\natexlab{b}}, J. Quant. Spectrosc. Radiat. Transf.,
  189, 267

\end{thebibliography}

\end{document}